\documentclass[twocolumn,showpacs,prd]{revtex4}

\newcommand{\beq}{\begin{equation}}
\newcommand{\eeq}{\end{equation}}
\newcommand{\bea}{\begin{eqnarray}}
\newcommand{\eea}{\end{eqnarray}}

\usepackage{graphicx}
\usepackage{amssymb}
\usepackage{amsmath}

\usepackage{psfrag}
\usepackage[final]{epsfig}

\begin{document}

\title{A single-domain spectral method for black hole puncture data}

\author{Marcus Ansorg, Bernd Br\"ugmann, Wolfgang Tichy}

\affiliation{
Center for Gravitational Physics and Geometry and
Center for Gravitational Wave Physics\\
Penn State University, University Park, PA 16802}

\date{June 25, 2004}

\begin{abstract}
We calculate puncture initial data corresponding to both single
and binary black hole solutions of the constraint equations by
means of a pseudo-spectral method applied in a single spatial
domain. Introducing appropriate coordinates, these methods exhibit
rapid convergence of the conformal factor and lead to highly
accurate solutions. As an application we investigate small mass
ratios of binary black holes and compare these with the
corresponding test mass limit that we obtain through a
semi-analytical limiting procedure. In particular, we compare the
binding energy of puncture data in this limit with that of a test
particle in the Schwarzschild spacetime and find that it deviates
by $50\%$ from the Schwarzschild result at the innermost stable
circular orbit of Schwarzschild, if the ADM mass at each puncture
is used to define the local black hole masses.
\end{abstract}

\pacs{
04.25.Dm, 
04.20.Ex, 
95.30.Sf    
%
\quad Preprint number: CGPG-03/12-3
}

\maketitle

\section{Introduction}

The evolution problem of general relativity requires the
specification of initial data that satisfies the Hamiltonian and
momentum constraints on the initial hypersurface. There are
different strategies to pose initial data for a specific physical
situation, which typically involve a choice of free data and the
subsequent numerical solution of the constraint equations to
obtain the physical data; see~\cite{Cook00a} for a recent review.
An active area of research is concerned with initial data that
describe two orbiting black holes
\cite{Cook93,Brandt97b,Matzner98a,Pfeiffer:2002xz,Tichy03a}. For
example, one can study the two-body problem of relativity by
constructing sequences of quasi-circular binary data sets that
describe the quasi-adiabatic inspiral of two black holes
\cite{Cook94,Baumgarte00a,Pfeiffer:2000um,Cook:2001wi,Gourgoulhon02,Grandclement02,Tichy02,Tichy03b}.
Furthermore, binary black hole data sets are the starting point
for evolutions in numerical relativity, e.g.\ \cite{Bruegmann:2003aw}.

Important aspects of black hole data sets are the choice of
hypersurface and how the physical singularity inside the black
holes is treated. Concretely, since the constraints give rise to
elliptic equations, one has to specify a computational domain and
boundary conditions. One possibility when considering two black
holes is to work on $\mathbb{R}^3$ with two balls excised. At the
spherical excision boundary one can impose boundary conditions
based on an isometry, as suggested by Misner~\cite{Misner60}. This
boundary condition is used in the first fully 3D numerical data
sets \cite{Cook93} and in the more recent thin sandwich type
initial data sets \cite{Gourgoulhon02}. The excision boundary can also
be defined by an apparent horizon boundary condition 
\cite{Thornburg85,Thornburg87}, see \cite{Cook:2001wi,Pfeiffer:2002wt}
for recent applications.  Other boundary conditions are motivated by
Kerr-Schild coordinates \cite{Matzner98a}. 

Excising spheres introduces a technical complication into
numerical methods on Cartesian grids. In finite differencing codes
on Cartesian grids, the boundary points are not aligned with the
grid and one has to construct appropriate stencils for a `lego'
sphere \cite{Cook93,Diener99,Hawley:2003az}. Alternatively, one
can work with adapted coordinates which match the spherical
boundary, for example \v{C}ade\v{z} coordinates \cite{Cook93}, or
one can use multiple coordinate patches with spherical coordinates
at the excision region
\cite{Thornburg85,Thornburg87,Gourgoulhon02,Grandclement02,Pfeiffer:2002wt}.

An alternative to excision boundaries is to work on $\mathbb{R}^3$
with two points (the `punctures') excised, where the punctures
represent the inner asymptotically flat infinity (Brill-Lindquist
topology~\cite{Brill63,Misner57}). Using the Brill-Lindquist topology
directly is problematic numerically since one has to resolve a one
over radius coordinate singularity. However, it is possible to
analytically compactify the inner asymptotically flat region, filling
in the missing puncture points, and to work on $\mathbb{R}^3$
\cite{Brandt97b,Beig94,Beig96,Beig:2000ei,Dain00}. This simplifies
the numerical method because no special inner boundary condition
has to be considered \cite{Brandt97b}.

In this paper we focus on the construction of an efficient numerical
method for the computation of black hole puncture data for vacuum
spacetimes containing one or two black holes with linear momentum and
spin. The numerical method, pseudo-spectral collocation, e.g.\
\cite{Boyd00}, can give exponential convergence when the solution is
infinitely often differentiable (${\cal C}^\infty$). However, in its
usual form puncture data is only ${\cal C}^2$ at the punctures. We
resolve this issue by constructing an appropriate coordinate
transformation that renders the puncture data smooth at the location
of the punctures. Consequently, our pseudo-spectral method converges
rapidly to highly accurate solutions, although the convergence rate is
generally not exponential due to logarithmic terms in expansions at
infinity, see below.

Note that spectral methods have already been applied successfully
to various elliptic problems in numerical relativity, including
neutron star initial data
\cite{Bonazzola98a,Bonazzola98b,Ansorg01b,Ansorg:2003br},
homogeneous star models \cite{Schobel:2003jg,Ansorg:2004vv},
relativistic Dyson rings \cite{Ansorg:2002vh}, and black hole
initial data
\cite{Gourgoulhon02,Grandclement02,Pfeiffer:2002xz,Pfeiffer:2002wt}.
In particular, \cite{Gourgoulhon02,Pfeiffer:2002wt} use several
coordinate patches to cover a binary black hole excision domain,
and the situation can become quite complicated with 43 rectangular
boxes and 3 spherical shells with various overlap and matching
boundary conditions \cite{Pfeiffer:2002wt}.  While such
multi-patch codes have a certain grid adaptivity built in (cmp.\
`spectral elements' \cite{Boyd00}), one of our motivations was to
simplify the spectral method by construction of a simpler
computational domain.

Therefore, a noteworthy feature of our spectral puncture method is
that our choice of coordinates maps $\mathbb{R}^3$, including
spatial infinity and the two puncture points, to a {\em single}
rectangular coordinate patch. This is the case for spherical
coordinates with a radial compactification, but recall that in
addition we want to ensure smoothness at the punctures.

The paper is organized as follows. In Sec.~\ref{SpMeth} we
describe our spectral method for the solution of the Hamiltonian
constraint on a single domain. After we introduce the puncture
data in Sec.~\ref{PuncData}, Sec.~\ref{OnePunc} discusses
analytical issues and numerical results for a single puncture. In
Sec.~\ref{TwoPuncs} we develop our single-domain spectral method
for two punctures and present the key result, which is rapid
convergence of this scheme to highly accurate solutions. 
The application of our method to the case of small mass ratios and a comparison with a
semi-analytic test mass limit can be found in
Sec.~\ref{TestMassLimit}. Finally, in Sec.~\ref{BindingEnergy} we
compute binding energies of puncture data in this limit. We
conclude in Sec.~\ref{Conclusion}.

\section{The spectral method}\label{SpMeth}

As will be discussed in detail in the subsequent sections, for
both the single and the two-puncture initial data problems an
elliptic equation of the form
\begin{equation}\label{elliptic}
    f(u)\equiv\triangle u +\varrho(u) = 0
\end{equation}
arises
for a function $u$. Here, $\triangle$ denotes the Laplace operator, and
$\varrho$ is a source term which in general depends on $u$.

In what follows we will introduce coordinates $(A,B,\varphi)$ with
\begin{equation}A\in[0,1], \quad B\in[-1,1],\quad\varphi\in[0,2\pi),\end{equation}
for each specific case that we consider, in which $u$ is well-defined
within the spatial domain, in particular at its boundaries.

As will become clear below, $u$ always obeys a physical fall-off condition
at spatial infinity,
\begin{equation}\lim_{r\to\infty} u=0.\end{equation}
Since in all cases to be considered, the coordinate $A$ is introduced such that
\begin{equation}r\to\infty\Longleftrightarrow A\to 1,\label{Aeq1}\end{equation}
we consider an additional function $U$ which is given by
\begin{equation}u=(A-1)U.\label{Uboundary}\end{equation}

In our spectral method, the values of this function $U$ are calculated
at the grid points $(A_i,B_j,\varphi_k)$, i.e.\
\begin{equation}U_{ijk}=U(A_i,B_j,\varphi_k),\end{equation}
\begin{equation} 0\leq i<n_A,\quad 0\leq j<n_B,\quad 0\leq k<n_\varphi,\end{equation}
where we choose
\begin{eqnarray}
  A_i&=&\sin^2\left[\frac{\pi}{2n_A}\left(i+\frac{1}{2}\right)\right] , \\
  B_j&=&-\cos\left[\frac{\pi}{n_B}\left(j+\frac{1}{2}\right)\right] , \\
  \varphi_k&=&2\pi\frac{k}{n_\varphi}.
\end{eqnarray}
Hence, the grid points $A_i$ and $B_j$ are the zeros of the Chebyshev
polynomials $T_{n_A}(1-2x)$ and $T_{n_B}(-x)$, respectively, whereas
the $\varphi_k$ represent the zeros of $\sin(n_\varphi\varphi)$. Our
spectral expansion is thus a Chebyshev expansion with respect to the
coordinates $A$ and $B$, and a Fourier expansion with respect to
$\varphi$.

The spectral method enables us to calculate first and second
derivatives of $U$ from the values $U_{ijk}$ at the above grid points
within the chosen approximation order which is given by the numbers
$(n_A,n_B,n_\varphi)$. Thus, for a vector
\begin{equation}
  \vec{U}=(U_{000},\ldots,U_{(n_A-1)(n_B-1)(n_\varphi-1)})^{T}
\end{equation}
we may fill another vector
\begin{equation}
  \vec{f}(\vec{U})=(f_{000},\ldots,f_{(n_A-1)(n_B-1)(n_\varphi-1)})^{T}
\end{equation}
by the evaluation of $f(u)$ at the grid points
$(A_i,B_j,\varphi_k)$. This results in a non-linear set of
simultaneous equations
\begin{equation}
    \vec{f}(\vec{U})=0
\label{nonlinsys}
\end{equation}
for the unknown $U_{ijk}$.

For the function $U$ we find particular boundary constraints by
considering the elliptic equation (\ref{elliptic}), written in
terms of $(A,B,\varphi)$ at $A=0$, $A=1$, $B=\pm1$. A solution $U$
that is regular with respect to $A$, $B$ and $\varphi$ must obey
these requirements, which therefore replace boundary conditions
that usually need to be imposed. These boundary constraints are
called `behavioral'~\cite{Boyd00}. In addition, a desired
$2\pi$-periodicity with respect to $\varphi$ is already `built-in'
by the particular choice of our basis functions. Accordingly,
using the spectral method with the above {\em interior}
collocation points, no further work with respect to the boundaries
needs to be done, for regularity and periodicity will be realized
automatically. Hence, there is no other requirement constraining
the function $U$. It is uniquely determined by the elliptic
equation (\ref{elliptic}).

For the numerical solution of the discrete equivalent, Eq.~(\ref{nonlinsys}),
we address its non-linearity by performing Newton-Raphson iterations.
The solution $\vec{U}$ is written as
\begin{eqnarray}
  \vec{U}&=&\lim_{N\to\infty} \vec{U}_N, \\
  \vec{U}_{N+1}&=&\vec{U}_N - \vec{V}_N,
\end{eqnarray}
where $\vec{V}_N$ satisfies the linear problem
\begin{equation}
\label{linsys}
  J_N \vec{V}_N = \vec{b}_N
\end{equation}
with
\begin{equation} J_N=\frac{\partial \vec{f}}{\partial\vec{U}}(\vec{U}_N),
\quad\vec{b}_N=\vec{f}(\vec{U}_N).\end{equation} There are
different ways to solve the linear system arising from
multi-dimensional spectral methods, although some effort has to be
made to obtain an efficient method since the one-dimensional
spectral differentiation matrices are not sparse and the
conditioning of the system can be problematic; see for example
\cite{Boyd00,Pfeiffer:2002wt}. We solve (\ref{linsys}) with the
preconditioned `Biconjugate Gradient Stabilized (BICSTAB)' method
\cite{Barrett93}, and the choice of preconditioner is crucial for
the overall efficiency of the method.

We construct a preconditioner which is based on a second order finite
difference representation of $J_N$. To this end we consider the linearized
differential equation corresponding to (\ref{elliptic}) on the equidistant
grid in coordinates
$(\alpha,\beta,\varphi)$ with
\begin{equation}  A = \sin^2\alpha,\quad B=-\cos\beta . \end{equation}
Apart from the uniform distribution of our grid points, these
coordinates have the additional advantage that $U$ becomes
symmetric with respect to the planes $\alpha=0$, $\alpha=\pi/2$,
$\beta=0$ and $\beta=\pi$. Therefore, it is possible to calculate
second order finite differencing approximations of first and
second derivatives at any grid point by taking into account
adjacent neighboring points only.

The resulting matrix has at most seven non-vanishing entries per
row and is therefore well suited for the application of a sparse
system solver. We use the program package `hypre' which offers a
variety of sparse matrix methods \cite{hypre_web}. A choice that
works well in this context is the `Generalized Minimal Residual
(GMRES)' method preconditioned with the algebraic multigrid code
`BoomerAMG' \cite{hypre_web}.

Our implementation of the above procedure uses the BAM code as
infrastructure \cite{Bruegmann:2003aw}. Although both BAM and hypre
support parallelization, we have not parallelized the elliptic
solves for our spectral method. Computation of the binary black
hole solution shown in Fig.~\ref{figbam} takes four minutes on a
Xeon/Linux workstation.

In what follows we evaluate the convergence of the spectral method by
computing the `global relative accuracy' defined by
\begin{equation}
  \delta_{n,m}(U)=\mbox{max}_{(A,B,\varphi)}|1-U_n/U_m|,
\label{accuracy}
\end{equation}
where $U_n$ denotes a specific $n$th order spectral approximation of
the function $U$. The maximum is typically evaluated over a regular
grid of $6^3$ points. We take $n_A=n_B=2n_\varphi=n$, with the only
exception $n_\varphi=4$ which we use in axisymmetric situations.

Choosing a large value $m$ defines a reference solution to which solutions at a
lower order of approximation $n < m$ can be compared. This method
gives a reliable characterization of convergence in our examples.
Furthermore, we reduce the error in the solution of the discrete non-linear system
(\ref{nonlinsys}) below the error due to the finite order of the spectral approximation
which therefore dominates the accuracy of the method.

The convergence rate of a spectral method is called {\em
exponential} if the logarithm of the total error of an approximate
solution depends linearly on the corresponding approximation order for
sufficiently large order. This behavior is usually encountered if the
underlying solution to be approximated is analytic everywhere on the
spectral domain. However, if the solution is only
${\cal C}^k$-differentiable, the logarithm of the total error depends
linearly on the logarithm of the approximation order. In particular,
the slope of this line is $(k+2)$, and the scheme is called {\em
algebraically} convergent to $(k+2)$-th-order. Hence, from the
numerical convergence of the spectral method one can deduce the
differentiability of the solution to be approximated (see Figs.\ 1, 2
and 4 below for representative examples corresponding to puncture initial
data).

\section{Puncture data}\label{PuncData}

In the ADM-formulation of a `3+1'-splitting of the spacetime
manifold, the vacuum Hamiltonian and the momentum constraint
equations of general relativity read as follows:
\begin{eqnarray}
{R}^2+K^2-K_{ij}K^{ij} &=& 0 , \\
{\nabla}(K^{ij}-\gamma^{ij}K)&=&0 .
\end{eqnarray}
Here $\gamma_{ij}$ is the 3-metric, $K_{ij}$ the extrinsic curvature,
$K$ its trace, and ${R},{\nabla}$ are the Ricci scalar and the
covariant derivative, respectively, associated with $\gamma_{ij}$.

Following York's conformal-transverse-traceless decomposition method
\cite{Cook00a}, we make the following assumptions for the metric and the
extrinsic curvature ($\delta_{ij}$ denotes the three-dimensional Kronecker symbol):
\begin{eqnarray}
   \gamma_{ij}&=&\psi^4\delta_{ij}, \\
   K_{ij}
     &=&\psi^{-2}\left(V_{j,i}+V_{i,j}-\frac{2}{3}\,
     \delta_{ij}\,\mathrm{div}\boldsymbol{V}\right) .
\end{eqnarray}
The initial data described by this method are conformally flat and
maximally sliced, $K = 0$. With this ansatz the Hamiltonian constraint
yields an equation for the conformal factor $\psi$,
\begin{equation}
    \label{Deltapsi}
  \triangle\psi+\frac{1}{8}\;\psi^5 K_{ij}K^{ij}=0,
\end{equation}
while the momentum constraint yields an equation for the vector
potential $\boldsymbol{V}$,
\begin{equation}
    \label{DeltaV}
   \triangle\boldsymbol{V}+\frac{1}{3}\;\mathrm{grad}(\mathrm{div}\boldsymbol{V})=0.
\end{equation}

One can proceed by choosing a non-trivial analytic solution of the
Bowen-York type for the momentum constraint,
\beq
  \boldsymbol{V}= \sum_{n=1}^{N_{p}} \left(
                  -\frac{7}{4|\boldsymbol{x}_n|}\boldsymbol{P}_n
          -\frac{\boldsymbol{x}_n\boldsymbol{\cdot P}_n}{4|\boldsymbol{x}_n|^3}\boldsymbol{\;x}_n
                    +\frac{1}{|\boldsymbol{x}_n|^3} \;\boldsymbol{x}_n\boldsymbol{\times S}_n \right),
\label{bowenyork} \eeq with poles at a finite number of $N_p$
spatial points, the locations of the punctures. Here the vector
parameters $\boldsymbol{P}_n$ and $\boldsymbol{S}_n$ can be
identified with the physical linear and angular momenta of the
$n$th puncture. The vector $\boldsymbol{x}_n$ points from the
$n$th puncture to the point $(x,y,z)$, $\boldsymbol{x}_n = (x-x_n,
y-y_n, z-z_n)^T$, and $|\boldsymbol{x}_n|$ is its Euclidian norm.

In \cite{Brandt97b} it is pointed out that for the extrinsic curvature
determined by (\ref{bowenyork}) a particular solution of the
Hamiltonian constraint is obtained by writing the conformal factor
$\psi$ as a sum of a singular term and a finite correction $u$,
\begin{equation}
\label{psi}
  \psi= 1+\sum_{n=1}^{N_p}\frac{m_n}{2|\boldsymbol{x}_n|}\;+\;u ,
\end{equation}
with $u\to0$ as $|\boldsymbol{x}_n|\to\infty$. The parameter $m_n$
is called the bare mass of the $n$th puncture.

The main point of the puncture construction is that in terms of $u$
the Hamiltonian constraint becomes a well-defined equation on the
entire Cartesian 3-space (see \cite{Dain01a} for a general existence
theorem for such asymptotically flat initial data). However, it turns
out that $u$ is in general only ${\cal C}^2$ at the punctures,
although it is ${\cal C}^\infty$ elsewhere.

As discussed in Section \ref{SpMeth}, such a drop of differentiability
implies that a spectral method can only be expected to be algebraically convergent
to fourth order. We first show for a single puncture that a simple coordinate 
transformation can resolve the differentiability problem at the location of 
the punctures. After that we discuss similar techniques for two punctures.

Note that by virtue of Theorem 1 by Dain and Friedrich \cite{Dain01a}, the conformal factor 
can only be expected to be globally $\cal{C}^\infty$-differentiable with respect to our 
coordinates $(A,B,\varphi)$ if the individual linear momenta $\boldsymbol{P}_n$ vanish. 
In fact, we will find that for punctures with linear momenta the conformal factor possesses
logarithmic terms when expanded at infinity, i.e.\ at $A=1$. This holds also
true if the total linear momentum, i.e.\ the sum of all $\boldsymbol{P}_n$ vanishes. 
Consequently, our single-domain spectral method cannot be exponentially convergent. 
Nevertheless, the scheme is rapidly converging towards highly accurate numerical solutions.

\section{Single-puncture initial data}
\label{OnePunc}

For a single puncture at the origin of a Cartesian grid, we introduce
spherical coordinates
$(r,\vartheta,\varphi)$ via
\begin{eqnarray}
  \nonumber x&=&r\cos\vartheta, \\
            y&=&r\sin\vartheta\cos\varphi, \\
  \nonumber z&=&r\sin\vartheta\sin\varphi,
\end{eqnarray}
where
\beq
  r\in[0,\infty),\quad \vartheta\in[0,\pi],\quad\varphi\in[0,2\pi).
\eeq
The conformal factor for a single puncture is
\begin{equation}
  \psi =1+\frac{m}{2r}+u,
\end{equation}
and we therefore choose to compactify the spatial domain
by introducing a new radial coordinate $A$ by
\begin{equation}
  A = \left(1+\frac{m}{2r}\right)^{-1},
\end{equation}
which implies (\ref{Aeq1}).

We separately investigate the two situations in which either the
linear momentum $\boldsymbol{P}$ or the spin $\boldsymbol{S}$
vanishes. A single black hole with either small Bowen-York spin or
linear momentum has also been considered in
\cite{Gleiser:1998ng,Gleiser:1999hw,Laguna:2003sr}.

\subsection{Single puncture with spin}

Consider first a single puncture with $\boldsymbol{P}=0$ and $S^i=S_x\delta^i_1$.
In the chosen spherical coordinates the Hamiltonian constraint reads
\begin{equation}
  \triangle\psi+\frac{9S_x^2}{16r^6}\psi^{-7}\sin^2\vartheta=0.
\end{equation}
For the auxiliary function $u$ we obtain a non-linear Poisson-like equation,
\begin{eqnarray}
  && u_{AA}+\frac{2u_A}{A}+\frac{1}{A^2(1-A)^2}
  \left(u_{\vartheta\vartheta}+u_\vartheta\cot\vartheta
  +\frac{u_{\varphi\varphi}}{\sin^2\vartheta}\right)
\nonumber \\
  && \qquad=-\frac{36w^2A(1-A)^2}{(1+Au)^7}\sin^2\vartheta
\end{eqnarray}
with $w=S_x/m^2$. The solution $u$ is uniquely determined by
regularity and periodicity conditions at $\vartheta=0$,
$\vartheta=\pi$ and $\varphi=0$, $\varphi=2\pi$, respectively.
For $A=0$ only a regularity condition needs to be imposed, while
for $A=1$ we set $u=0$. Thus, the single-domain spectral method
described in Sec.~\ref{SpMeth} is applicable with
\begin{equation}B=2\vartheta/\pi-1,\end{equation}
provided that a global regular solution exists.

In order to study the behavior of $u$ globally, and in particular
close to the puncture, consider the following Taylor series which
converges for sufficiently small $w$:
\begin{equation}
  u=\sum_{j=1}^\infty w^{2j}u_j.
\end{equation}
All $u_j$ can explicitly be given in closed analytic form.
In particular, for $u_1$ we obtain ($P_2$ denotes the second Legendre polynomial):
\begin{eqnarray}
  u_1 &=& u_{1,0}+u_{1,2}P_2(\cos\vartheta) ,\\ 
  u_{1,0}&=& \frac{2}{5}(-2A^5+6A^4-5A^3+1), \\ 
  u_{1,2}&=& \frac{4}{5}(1-A)^3A^2.
\end{eqnarray}
Note that $u_1$ is regular at $A=0$ in the spherical
coordinates $(A,\vartheta,\varphi)$. The same holds for all $u_j$ and
in fact for $u$, see \cite{Dain01a}. Hence, the ${\cal C}^2$-differentiability of $u$ at
the puncture has been translated into a ${\cal C}^\infty$-differentiability with
respect to spherical coordinates. We can still recognize the
original behavior of the function which is exhibited by the fact that
$u_1$ possesses a term $\sim r^3$ which is ${\cal C}^2$-differentiable
in Cartesian coordinates. However, in the chosen spherical coordinates
$(A,\vartheta,\varphi)$, $u$ becomes globally ${\cal C}^\infty$. 

\begin{figure}[h]
    \unitlength1cm
    \begin{picture}(7,6)
        \psfrag{x}[t][r]{{$N$}}
        \psfrag{deltaU}[c][c]{{$\delta_{n,50}(U)$}}
        \psfrag{x}[c][c]{{$n$}}
        \psfrag{x1}[t][c]{{\small $10$}}
        \psfrag{x2}[t][c]{{\small $15$}}
        \psfrag{x3}[t][c]{{\small $20$}}
        \psfrag{x4}[t][c]{{\small $25$}}
        \psfrag{x5}[t][c]{{\small $30$}}
        \psfrag{y1}[r][c]{{\small $10^{-3}$}}
        \psfrag{y2}[r][c]{{\small $10^{-5}$}}
        \psfrag{y3}[r][c]{{\small $10^{-7}$}}
        \psfrag{y4}[r][c]{{\small $10^{-9}$}}
        \psfrag{y5}[r][c]{{\small $10^{-11}$}}
        \psfrag{y6}[r][c]{{\small $10^{-13}$}}
        \put(0,0){\epsfig{file=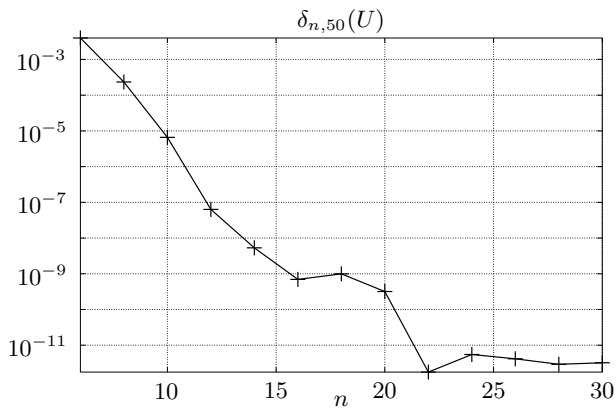,scale=0.6}}
    \end{picture}
    \caption{For a single puncture with vanishing linear momentum
    parameter the spin $S^i=m^2 w\,\delta_1^i$ with $w=0.2$ has
    been chosen. The plot shows the relative global accuracy of
    the spectral method for expansion order $n_A=n_B=n$
    compared to a reference solution with $n = 50$, see (\ref{accuracy}).
    For this axisymmetric example we have used $n_\varphi=4$.}
\label{figspin}
\end{figure}

Consequently, the application of our single-domain spectral method exhibits
exponential convergence, which can be seen in Fig.\ \ref{figspin} for a
representative example.

\subsection{Single puncture with linear momentum}
\label{onepuncwithP}

Consider now a single puncture with linear momentum
$P^i=P_x\delta^i_1$ and vanishing spin. The Hamiltonian constraint becomes
\begin{equation}
  \triangle\psi+\frac{9P_x^2}{16r^4}\psi^{-7}(1+2\cos^2\vartheta)=0.
\end{equation}
Similar to the treatment in the previous section we obtain in
spherical coordinates the non-linear Poisson-like equation
\begin{eqnarray}
  && u_{AA}+\frac{2u_A}{A}+\frac{1}{A^2(1-A)^2}
  \left(u_{\vartheta\vartheta}+u_\vartheta\cot\vartheta
  +\frac{u_{\varphi\varphi}}{\sin^2\vartheta}\right)
\nonumber \\
  &&\qquad =-\frac{9v^2A^3}{4(1+Au)^7}(1+2\cos^2\vartheta)
\end{eqnarray}
with $v=P_x/m$. Again, we may study the behavior of $u$
by performing a Taylor expansion which converges for sufficiently small $v$,
\begin{equation}
  u=\sum_{j=1}^\infty v^{2j}u_j .
\end{equation}
All $u_j$ can explicitly be given in closed analytic form.
In particular, for $u_1$ we obtain
\begin{eqnarray}
  u_1 &=& u_{1,0}+u_{1,2}P_2(\cos\vartheta) ,\\ 
  u_{1,0}&=& \frac{1}{8}(1-A^5), \\ 
  u_{1,2}&=& \frac{(1-A)^2}{20A^3} [ 84(1-A)\log(1-A) \nonumber \\
         && \mbox{}+84A-42A^2-14A^3\nonumber \\&&\mbox{}-7A^4-4A^5-2A^6 ].
\end{eqnarray}
We recover that $u$ is analytic at $A=0$ while it is
${\cal C}^4$-differentiable in Cartesian coordinates, which is
implied by a term $\sim r^5$. 

\begin{figure}[h]
    \unitlength1cm
    \begin{picture}(7,6)
        \psfrag{deltaU}[c][c]{{$\delta_{n,70}(U)$}}
        \psfrag{x}[t][c]{{$\log_{10} n$}}
        \psfrag{x1}[t][c]{{\small $1$}}
        \psfrag{x2}[t][c]{{\small $1.2$}}
        \psfrag{x3}[t][c]{{\small $1.4$}}
        \psfrag{x4}[t][c]{{\small $1.6$}}
        \psfrag{y1}[r][c]{{\small $10^{-3}$}}
        \psfrag{y2}[r][c]{{\small $10^{-4}$}}
        \psfrag{y3}[r][c]{{\small $10^{-5}$}}
        \psfrag{y4}[r][c]{{\small $10^{-6}$}}
        \put(0,0){\epsfig{file=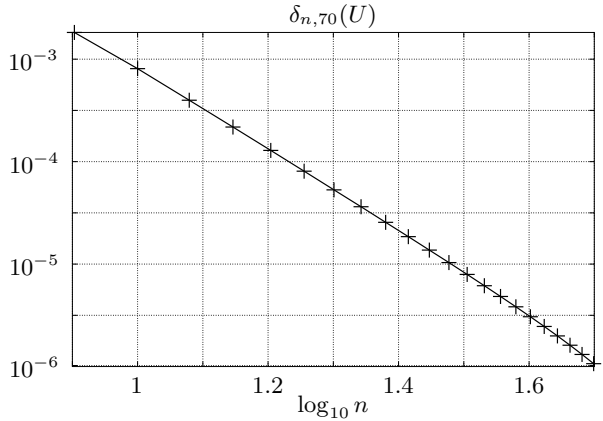,scale=0.6}}
    \end{picture}
    \caption{For a single puncture with vanishing spin parameter
    the linear momentum $P^i=mv\,\delta^i_1$ with $v=0.2$ has been chosen.
    The plot shows the relative global accuracy of
    the spectral method for expansion order $n_A=n_B=n$
    compared to a reference solution with $n = 70$, see (\ref{accuracy}).
    For this axisymmetric example we have used $n_\varphi=4$.}
\label{figmom}
\end{figure}

However, the solution $u=u(A,\vartheta,\varphi)$ also
possesses logarithmic terms with a branch point at $A=1$
($r\to\infty$).
For a single puncture, such logarithmic terms are known to occur 
for non-vanishing linear momentum, e.g.\ \cite{Gleiser:1999hw,Dain01a}.
In particular, the leading term
\begin{equation} \frac{21(1-A)^3}{5A^3} \log(1-A) \end{equation}
gives rise to a mere ${\cal C}^2$-differentiability of $u$ at
$A=1$. This fact again is reflected by the spectral
method, which now converges only algebraically to fourth order as
expected, see Fig.\ \ref{figmom} for a representative example.

\section{Two-puncture initial data}
\label{TwoPuncs}

\begin{figure*}[t]
    \unitlength1cm
    \begin{picture}(14,12)
        \psfrag{(a)}[b][c]{{$(a)$}}
        \psfrag{(b)}[b][c]{{$(b)$}}
        \psfrag{(c)}[b][c]{{$(c)$}}
        \psfrag{(d)}[b][c]{{$(d)$}}

        \psfrag{u}[c][r]{{$A$}}
        \psfrag{v}[c][r]{{$B$}}
        \psfrag{xlow}[t][r]{{\small $\;0$}}
        \psfrag{xmed}[c][c]{{\small $$}}
        \psfrag{xhigh}[t][r]{\small {$1$}}
        \psfrag{ylow}[r][r]{{\small $-1\;\,$}}
        \psfrag{ymed}[r][r]{{\small $0$}}
        \psfrag{yhigh}[r][r]{\small {$1\;\,$}}

        \put(0,6){\epsfig{file=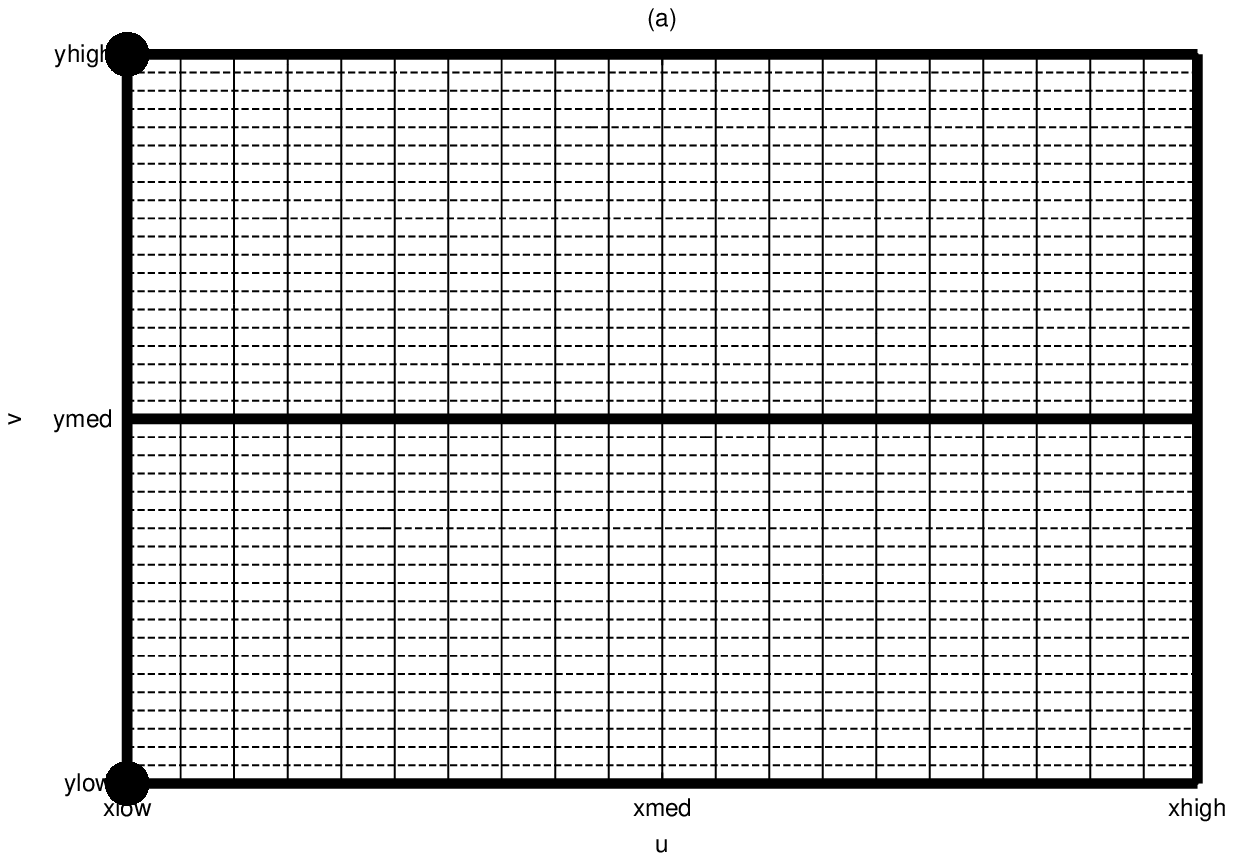,scale=0.5}}

        \psfrag{X}[c][r]{{$\xi$}}
        \psfrag{R}[c][r]{{$\eta$}}
        \psfrag{xlow}[t][r]{{\small $\;0$}}
        \psfrag{xmed}[t][c]{{\small $2$}}
        \psfrag{xhigh}[t][c]{\small {$4$}}
        \psfrag{ylow}[r][r]{{\small $0\;\,$}}
        \psfrag{ymed}[r][r]{{\small $\frac{\pi}{2}$}}
        \psfrag{yhigh}[r][r]{\small {$\pi\;\,$}}

        \put(0,0){\epsfig{file=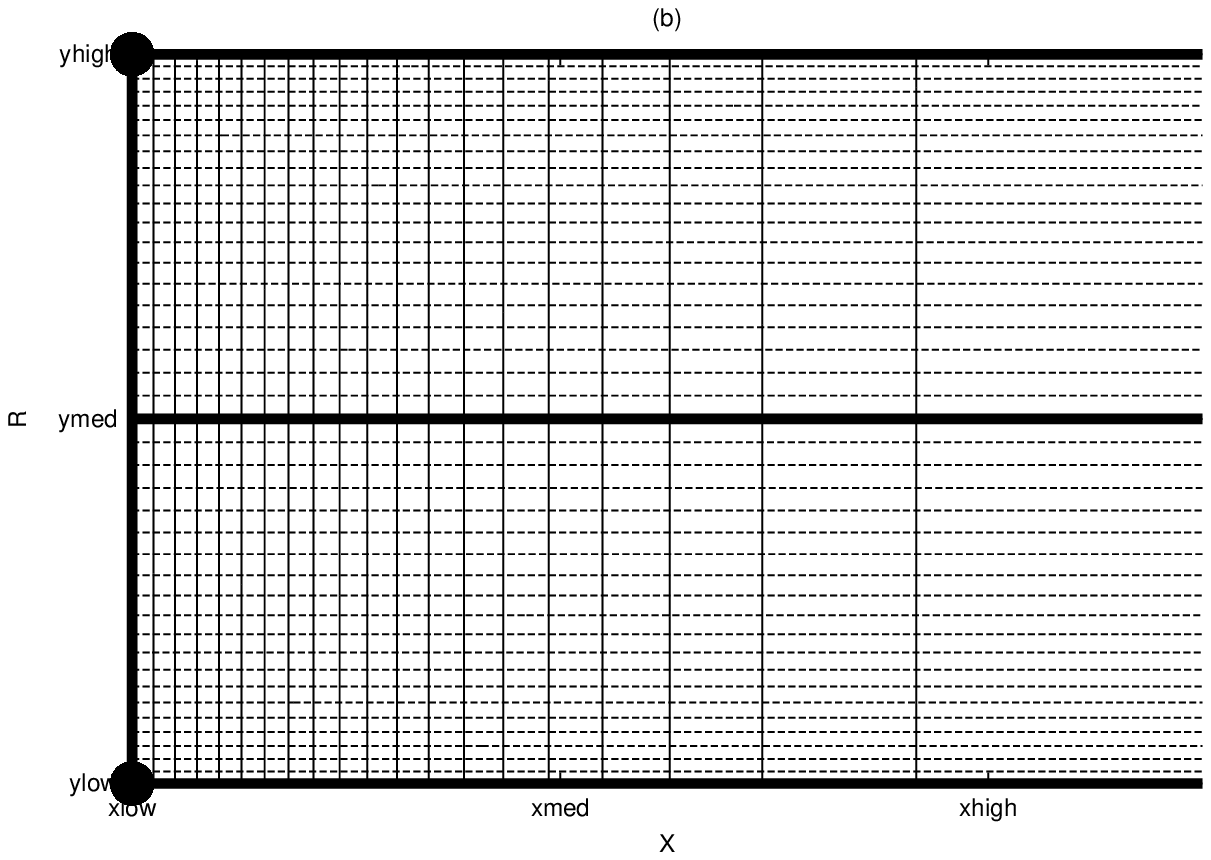,scale=0.5}}

        \psfrag{xi}[c][r]{{$X$}}
        \psfrag{rho}[c][r]{{$R$}}
        \psfrag{x1}[t][r]{{\small $-2$}}
        \psfrag{x2}[t][r]{{\small $-1$}}
        \psfrag{x3}[t][r]{{\small $$}}
        \psfrag{x4}[t][r]{{\small $\;\;\;1$}}
        \psfrag{x5}[t][r]{{\small $2$}}
        \psfrag{y1}[r][r]{{\small $0$}}
        \psfrag{y2}[r][r]{{\small $1$}}
        \psfrag{y3}[r][r]{{\small $2$}}
        \psfrag{y4}[r][r]{{\small $3$}}

        \put(7,6){\epsfig{file=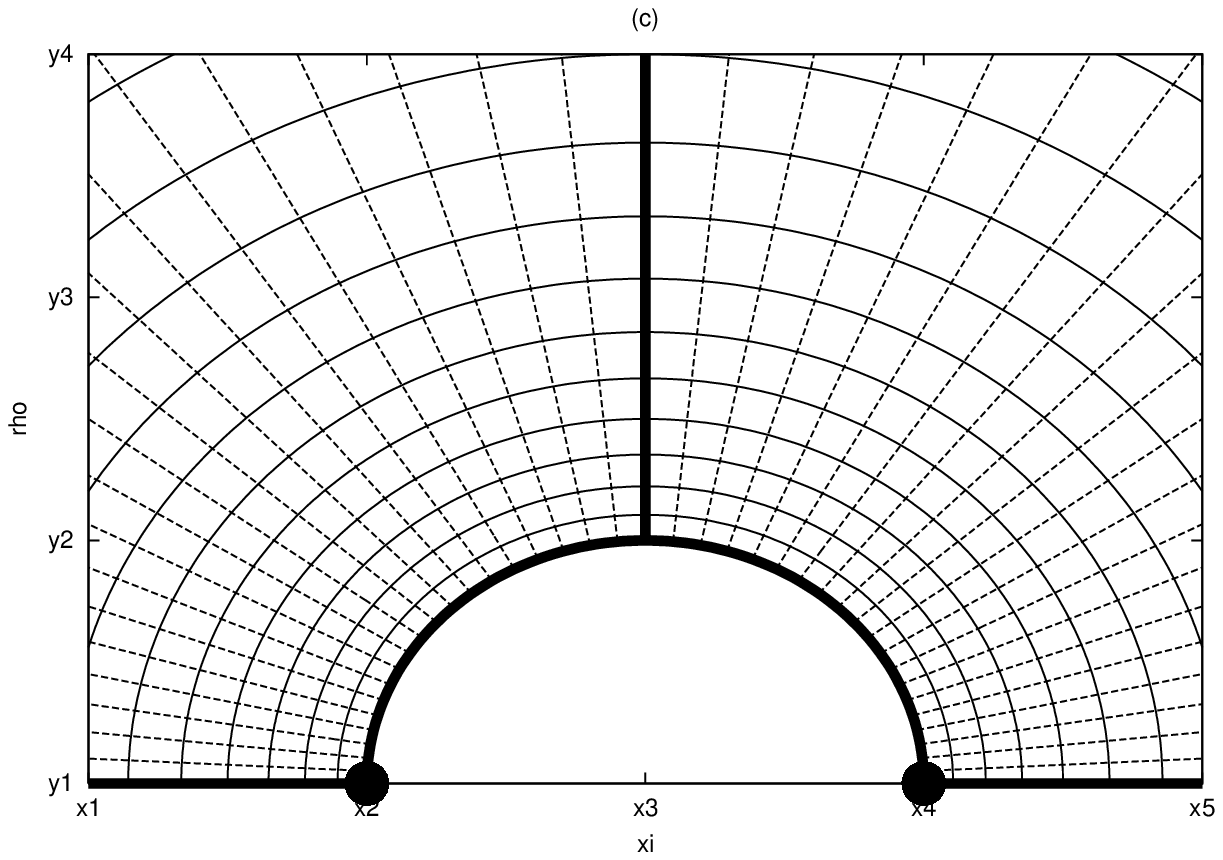,scale=0.5}}

        \psfrag{x/b}[c][r]{{$x/b$}}
        \psfrag{r/b}[c][r]{{$\rho/b$}}
        \psfrag{x1}[t][r]{{\small $-2$}}
        \psfrag{x2}[t][r]{{\small $-1$}}
        \psfrag{x3}[t][r]{{\small $$}}
        \psfrag{x4}[t][r]{{\small $\;\;\;1$}}
        \psfrag{x5}[t][r]{{\small $2$}}
        \psfrag{y1}[r][r]{{\small $0$}}
        \psfrag{y2}[r][r]{{\small $1$}}
        \psfrag{y3}[r][r]{{\small $2$}}
        \psfrag{y4}[r][r]{{\small $3$}}

        \put(7,0){\epsfig{file=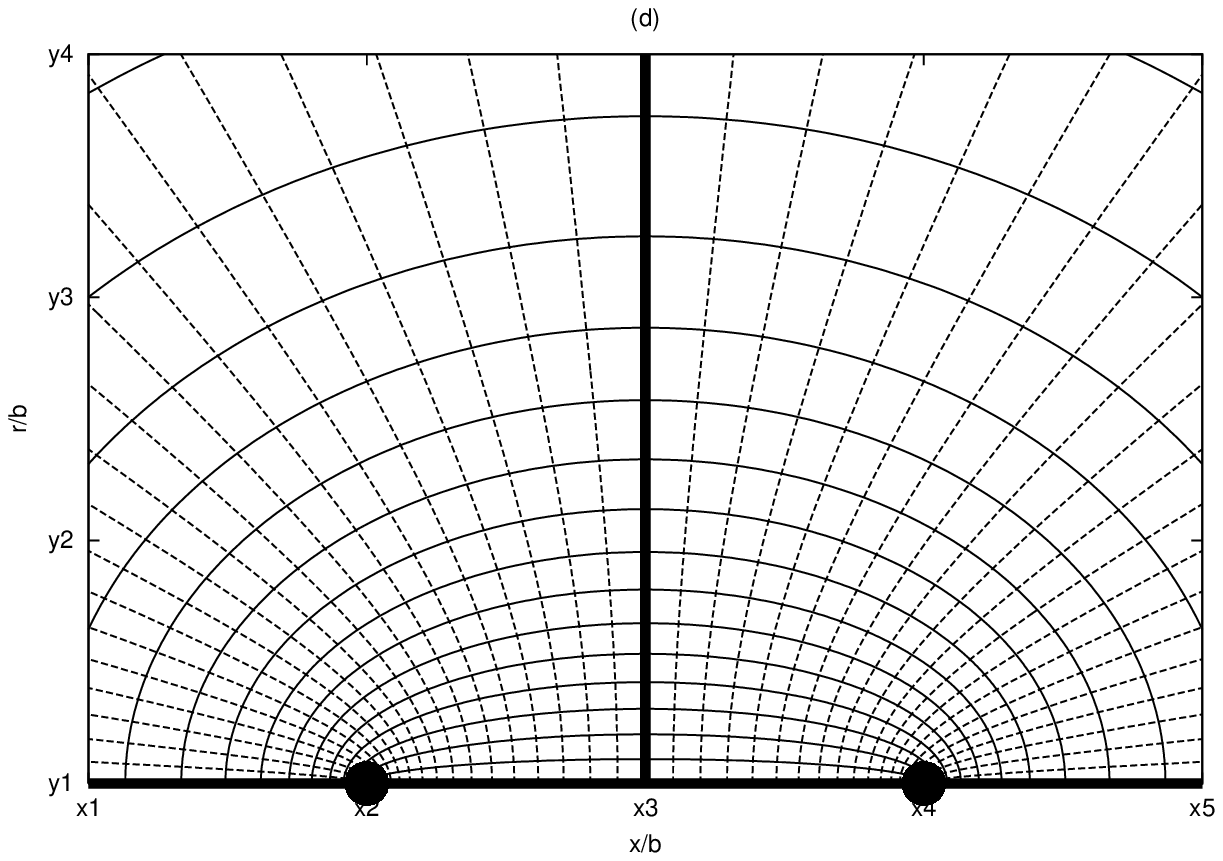,scale=0.5}}
        \end{picture} \caption{Several coordinate patches for
        the two puncture initial data problem. Shown are $(a)$
        equidistant coordinate lines in the system of spectral
        coordinates $(A,B)$, as well as $(b)$ their images in
        prolate spheroidal coordinates $(\xi,\eta)$, $(c)$ in
        the coordinates $(X,R)$, and $(d)$ in cylindrical
        coordinates $(x,\rho)$. The punctures are indicated by
        bullets. The $(x=0)$-plane, several sections of the
        $x$-axis and their corresponding images in the other
        coordinate systems as well as spatial infinity given
        by $A=1$ are emphasized by thick lines.}
\label{figcoords}
\end{figure*}

Consider two punctures that are placed symmetrically on the $x$-axis
at $x = \pm\, b$ so that $D=2b$ is the distance between the two
punctures.  We denote the bare mass of the punctures by $m_\pm$, the
linear momenta by $\boldsymbol{P}_\pm$ and the spin parameters by
$\boldsymbol{S}_\pm$; the subscripts refer to the corresponding
locations at $x=\pm b$.

In the following we introduce appropriate coordinates in which the
auxiliary function $u$ becomes regular at the location of the punctures.

The decomposition (\ref{psi}) reads
\begin{equation} \psi= 1+\frac{m_+}{2r_+}+\frac{m_-}{2r_-}\;+\;u , \end{equation}
with the distances from the punctures given by
\begin{equation} r_\pm = \sqrt{(x\mp b)^2+y^2+z^2} . \end{equation}

As we have seen for the single puncture initial data problem, the
auxiliary function $u$ discussed there is regular at the location of
the puncture in spherical coordinates about this point. We therefore
expect a similar regular behavior if we were to introduce coordinates
that become spherical at both punctures. However, regularity of $u$ at
the punctures can also be achieved if we use specific coordinates in which
the distances $r_\pm$ are analytic functions there (see \cite{Dain01a}). 
This is a weaker condition because it does not necessarily require one of our
coordinates to behave as $r_\pm$ close to the punctures. 

A coordinate transformation that describes this situation at the origin
in two dimensions is given by
\begin{equation}\label{cEQC2} c=C^2, \end{equation}
where
\begin{equation}\label{cAndC} c=x+\mbox{i}y \quad \mbox{and} \quad C=X+\mbox{i}Y \end{equation}
are complex combinations of Cartesian coordinates $(x,y)$ and new
coordinates $(X,Y)$.
Clearly, the distance becomes regular with respect to $X$ and $Y$,
\begin{equation}\label{distance}
  \sqrt{x^2+y^2} = \sqrt{c\bar{c}} = C\bar{C} = X^2+Y^2.
\end{equation}
Note that transformation (\ref{cEQC2}) maps a right angle at the origin to a
straight line through the origin.

For the two-puncture initial data problem, we apply this idea by
introducing a specific mapping
\begin{equation} (A,B,\varphi) \mapsto (x,y,z),\end{equation}
which is composed of several transformations (see Fig.\ \ref{figcoords}),
\begin{eqnarray}
(A,B,\varphi) & \mapsto & (\xi,\eta,\varphi) \mapsto (X,R,\varphi) \mapsto (x,\rho,\varphi)
\nonumber \\
&\mapsto& (x,y,z).
\end{eqnarray}
These transformations are chosen to realize the two different aspects
of the desired entire transformation, (i) regularity of $r_\pm$
at both punctures, and (ii)
mapping of a compact rectangular domain in $\mathbb{R}^3$
to the entire space of $(x,y,z)$-coordinates.

We first introduce cylindrical coordinates $(x,\rho,\varphi)$ such
that
\begin{equation} y = \rho\cos\varphi, \quad z=\rho\sin\varphi,
\quad\varphi\in[0,2\pi),\end{equation}
and combine $x$ and $\rho$ to form $c$,
\begin{equation} c = x+\mbox{i}\rho .\end{equation}
Now consider the transformation
\begin{equation} c = \frac{b}{2}\left(C+C^{-1}\right),
    \quad \mbox{where} \quad C = X+\mbox{i}R.
\end{equation}
It maps the region of the upper half plane with coordinates
$(X,R)$ which is exterior to the unit circle onto the upper half
plane of our coordinates $(x,\rho)$, see Fig.\
\ref{figcoords}$(c,d)$. \\
The key motivation behind this transformation has been to produce
locally at each puncture the same effect on angles as has been
done above in the transformation (\ref{cEQC2}), (\ref{cAndC}) and
which has resulted in the regular expression (\ref{distance}) for
the distance from the origin. Similarly we now obtain expressions
for the distances from either puncture
\begin{equation} r_\pm = |c\mp b|=\frac{b}{2\sqrt{X^2+R^2}}
            \left[\left(X\mp 1\right)^2+R^2\right]
\end{equation}
which are regular with respect to $X$ and $R$ at the punctures, i.e.\
at $c=\pm b$ or $C=\pm 1$.

Next we need to find a transformation which maps a compact rectangular region onto
the region of $(X,R)$-coordinates. As a first step, the polar transformation
\begin{equation} 
	C = \mbox{e}^\zeta,\quad 
	\zeta=\xi+\mbox{i}\eta, \quad \xi\in[0,\infty)\,,
	\eta\in[0,\pi]
\end{equation}
yields a strip which is infinitely extended with respect to positive $\xi$-values,
see Fig.\ \ref{figcoords}$(b)$.  Writing $c$ in terms of $\zeta$ gives
\begin{equation} c = b\cosh \zeta.\end{equation}
Thus we recover the transformation
\begin{equation} x = b\cosh \xi \cos \eta,\quad \rho=b\sinh\xi \sin\eta,\end{equation}
which maps the well-known prolate spheroidal coordinates
$(\xi,\eta)$ onto cylindrical coordinates. Hence, constant $\xi$-
and $\eta$-values correspond to confocal ellipses and hyperbolas,
respectively, in the $(x,\rho)$-plane. Their focal points are
located at the two punctures, i.e.\ at $(\xi,\eta)=(0,0)$ and
$(\xi,\eta)=(0,\pi)$, see Fig.\ \ref{figcoords}. The distances
$r_\pm$ expressed in terms of $\xi$ and $\eta$ are
\begin{equation} r_\pm = b(\cosh\,\xi \mp \cos \eta). \end{equation}
For a compactification we choose the relations
\begin{equation}
\xi = 2 \; \mbox{artanh\;}A\;,\quad \eta=\frac{\pi}{2}+2\arctan B .
\end{equation}
In summary, the transformation from $(A,B,\varphi)$ to
$(x,y,z)$ takes the (somewhat symmetric) form
\begin{eqnarray}
    \nonumber x&=&b\,\frac{A^2+1}{A^2-1}\,\frac{2B}{1+B^2}, \\
    y&=&b\,\frac{2A}{1-A^2}\,\frac{1-B^2}{1+B^2}\cos\varphi , \\
    \nonumber z&=&b\,\frac{2A}{1-A^2}\,\frac{1-B^2}{1+B^2}\sin\varphi .
\end{eqnarray}

It is now straightforward to apply our single-domain spectral
method to solve the Hamiltonian constraint (\ref{Deltapsi}) for
the two-puncture initial data problem. Again we impose $u\to 0$ as
$A\to 1$, i.e.\ $(x^2+y^2+z^2)\to\infty$. As in the one-puncture
initial data problem, at all the other boundaries we again merely
require regularity of the solution which replaces a particular
boundary condition there. As expected, the auxiliary function $u$
is $\cal{C}^\infty$ at the two punctures. 

As mentioned at the end of Section \ref{PuncData}, in general $u$
possesses logarithmic terms when expanded at infinity, $A=1$. In
\cite{Dain01a} a theorem is proved that does not exclude the existence of
such logarithmic terms given the fall-off condition satisfied by the
extrinsic curvature that we consider here. We have checked in the case
of axisymmetry analytically that for two equal mass punctures with
linear momentum logarithmic terms do indeed occur.
The only exception is when both linear momenta $\boldsymbol{P}_\pm$
vanish, in which case the solution is also $\cal{C}^\infty$ at $A=1$.
Otherwise we obtain terms $\sim (1-A)^3\log(1-A)$ if the total
momentum $\boldsymbol{P}=\boldsymbol{P}_+ + \boldsymbol{P}_-\neq 0 $,
and terms $\sim (1-A)^5\log(1-A)$ if $\boldsymbol{P}=0$.  In other
words, in the center of mass frame where the total linear momentum
vanishes the leading order logarithmic terms cancel, but next to
leading order logarithmic terms are still present such that the
solution is only ${\cal C}^4$ at $A=1$. Although we carried out this
analysis for axisymmetry, it is to be expected that the same result
applies to puncture data describing orbiting black holes in the center
of mass frame.

\begin{figure}[h]
    \unitlength1cm
    \begin{picture}(7,6)
        \psfrag{o}[r][c]{{Entire space ($6^3$ points)}}
        \psfrag{p}[r][c]{{Infinity ($6^2$ points)}}
        \psfrag{q}[r][c]{{Puncture ($6$ points)}}
        
		\psfrag{deltaU}[c][c]{{$\delta_{n,70}(U)$}}
		\psfrag{x}[t][c]{{$n$}} \psfrag{x1}[t][c]{{\small
		$10$}} \psfrag{x2}[t][c]{{\small $20$}}
		\psfrag{x3}[t][c]{{\small $40$}}
		\psfrag{x4}[t][c]{{\small $50$}}
		\psfrag{y1}[r][c]{{\small $10^{-3}$}}
		\psfrag{y2}[r][c]{{\small $10^{-5}$}}
		\psfrag{y3}[r][c]{{\small $10^{-7}$}}
		\psfrag{y4}[r][c]{{\small $10^{-9}$}}
		\psfrag{y5}[r][c]{{\small $10^{-11}$}}
		\put(0,0){\epsfig{file=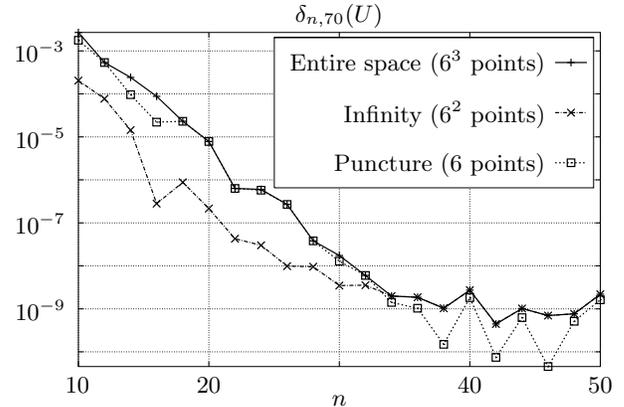,scale=0.6}}
		\end{picture} 
\caption{
Two punctures with vanishing spins. The physical parameters are given
by $m_+=m_-=b, P^i_\pm=\pm0.2\,b\,\delta^i_2$.  For this plot we took
$n_A=n_B=2n_\varphi=n$ and compared to a reference solution with $n =
70$. Apart from the global relative accuracy (see (\ref{accuracy}))
taken over $6^3$ spatial points, the corresponding maximal deviations
at infinity and at the punctures are shown. For small $n$, the error
near the punctures is about ten times larger than the error at
infinity, and the convergence rate is approximately exponential down
to about $10^{-9}$. The error at infinity converges at roughly sixth
algebraic order as expected, and for sufficiently large $n$ this
becomes the dominant convergence rate.  }
\label{figtwopunc}
\end{figure}

A representative convergence rate of our single-domain spectral method
is displayed in Fig.~\ref{figtwopunc}. We show the relative
accuracy (\ref{accuracy}), which involves the maximum over a set of
points, computed over a 3D set of points as before, but also for 
points only at infinity and at the puncture. The error is dominated by
errors near the puncture down to about $10^{-9}$ for $n < 35$. In this
regime convergence of the maximal error is exponential. However, the
error at infinity only converges algebraically at roughly sixth-order
as expected. Around $n=35$, the error at infinity overtakes the error
elsewhere and the overall convergence becomes algebraic.

Therefore we conclude that our numerical method is successful since it
obtains exponential convergence for orbiting punctures down to about
$10^{-9}$ with relatively small computational resources. At the
punctures our coordinate transformation leads to a smooth solution,
but at infinity there are logarithmic terms which lead to algebraic
convergence of sixth order. We consider this quite satisfactory since
higher accuracy is rarely needed in numerical relativity. In
principle, it should be possible to eliminate the leading logarithmic
term which should bring the calculation close to numerical round-off
errors. However, it is unclear whether logarithmic terms can be
avoided completely in this approach, for example by an appropriate
coordinate transformation.

\begin{figure*}[t]
    \unitlength1cm
    \begin{picture}(14,14)

        \psfrag{u}[c][r]{{$u$}}
        \psfrag{x}[c][t]{{$\begin{array}{cc} \\ x/M \\ (y=z=0)\end{array}$}}
        \psfrag{x1}[t][c]{{\small $$}}
        \psfrag{x2}[t][c]{{\small $-20$}}
        \psfrag{x3}[t][c]{\small {$$}}
        \psfrag{x4}[t][c]{{\small $$}}
        \psfrag{x5}[t][c]{{\small $$}}
        \psfrag{x6}[t][c]{\small {$20$}}
        \psfrag{x7}[t][c]{\small {$$}}
        \psfrag{y1}[r][r]{{\small $0$}}
        \psfrag{y2}[r][r]{{\small $0.005$}}
        \psfrag{y3}[r][r]{\small {$0.01$}}
        \psfrag{y4}[r][r]{{\small $0.015$}}

        \put(-1.5,7){\epsfig{file=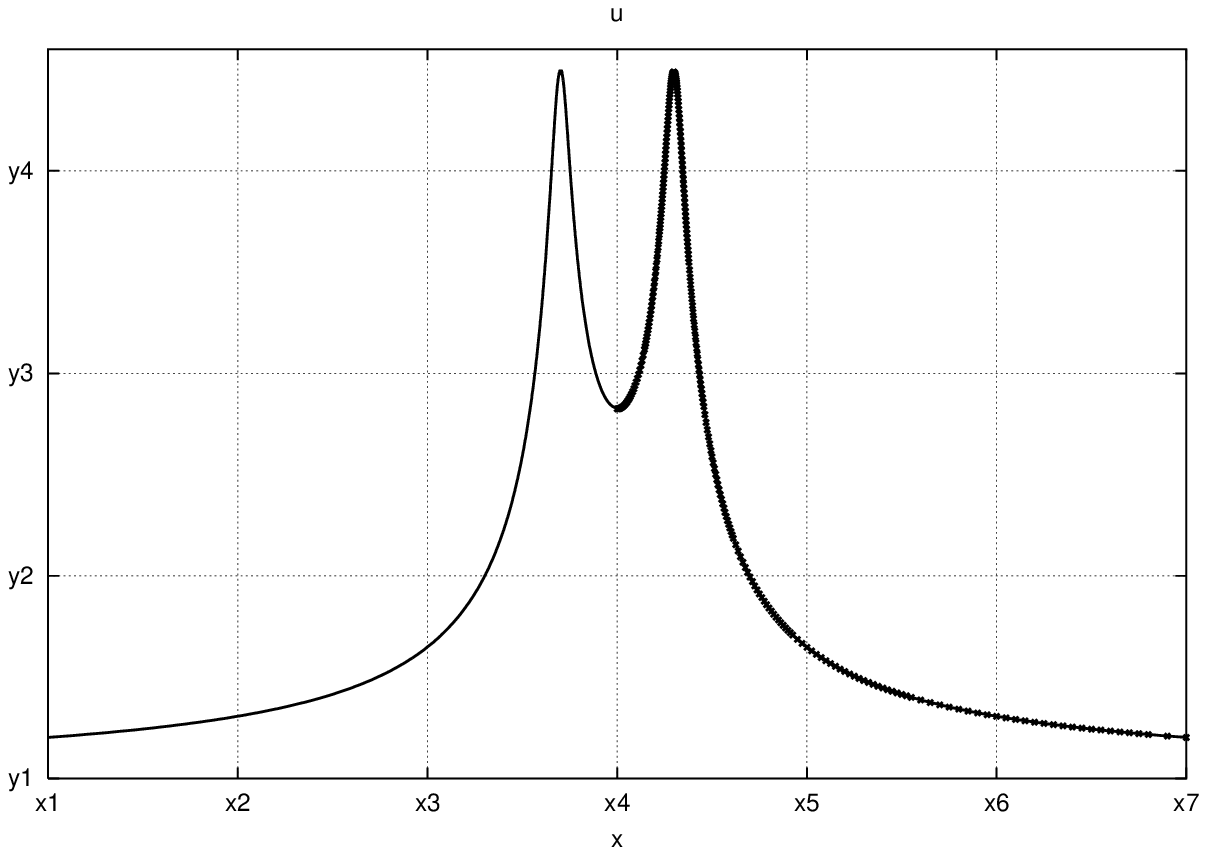,scale=0.6}}

        \psfrag{u}[c][r]{{$u$}}
        \psfrag{x}[c][t]{{$\begin{array}{cc} \\ x/M \\ (y=z=0)\end{array}$}}
        \psfrag{x1}[t][r]{{\small $2.7$}}
        \psfrag{x2}[t][r]{{\small $2.85$}}
        \psfrag{x3}[t][r]{{\small $$}}
        \psfrag{x4}[t][r]{{\small $3.15$}}
        \psfrag{x5}[t][r]{{\small $3.3$}}
        \psfrag{y1}[r][t]{{\small $0.016$}}
        \psfrag{y2}[r][r]{{\small $0.017$}}
        \psfrag{y3}[r][r]{{\small $0.018$}}

        \put(8,7){\epsfig{file=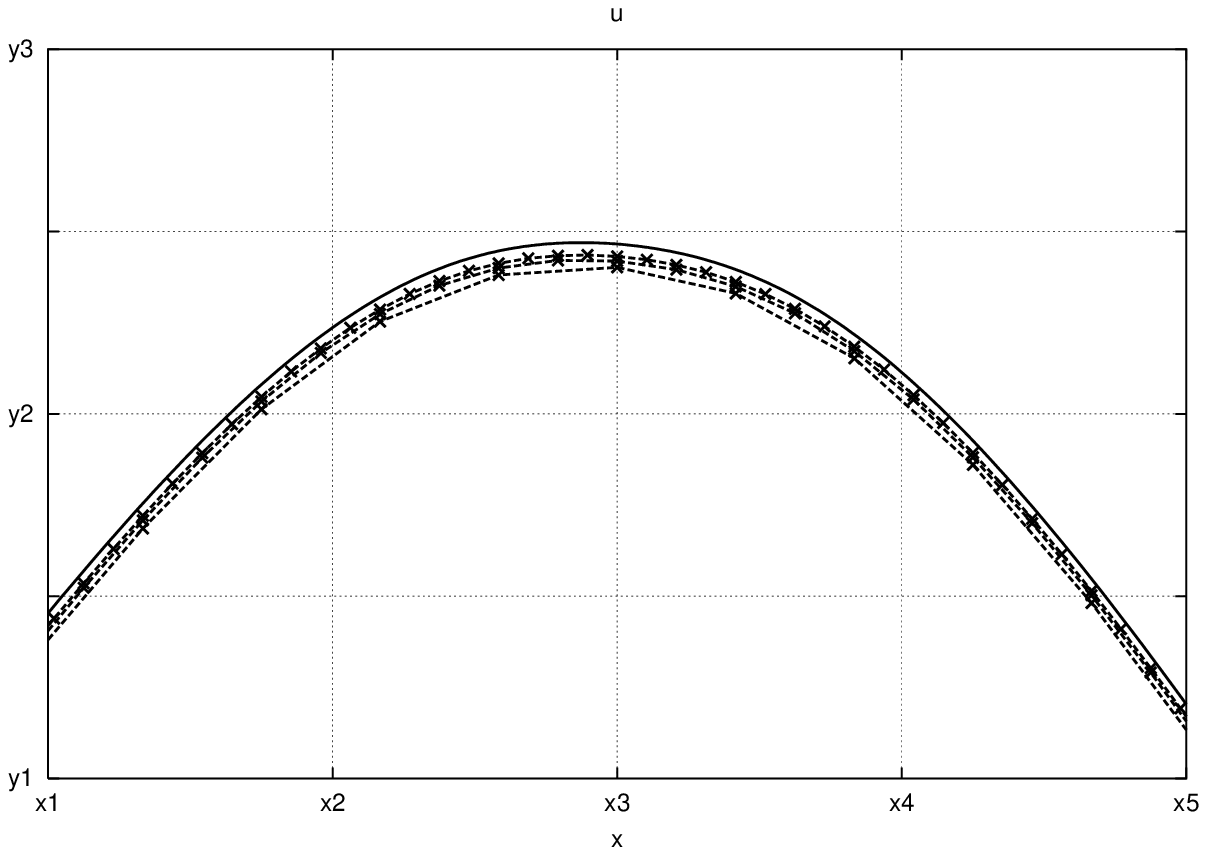,scale=0.6}}

        \psfrag{u}[c][r]{{$u$}}
        \psfrag{y}[c][t]{{$\begin{array}{cc} \\ y/M \\[1mm] (x=3M,\,z=0)\end{array}$}}
        \psfrag{x1}[t][c]{{\small $$}}
        \psfrag{x2}[t][c]{{\small $-20$}}
        \psfrag{x3}[t][c]{\small {$$}}
        \psfrag{x4}[t][c]{{\small $$}}
        \psfrag{x5}[t][c]{{\small $$}}
        \psfrag{x6}[t][c]{\small {$20$}}
        \psfrag{x7}[t][c]{\small {$$}}
        \psfrag{y1}[r][r]{{\small $0$}}
        \psfrag{y2}[r][r]{{\small $0.005$}}
        \psfrag{y3}[r][r]{\small {$0.01$}}
        \psfrag{y4}[r][r]{{\small $0.015$}}

        \put(-1.5,0.5){\epsfig{file=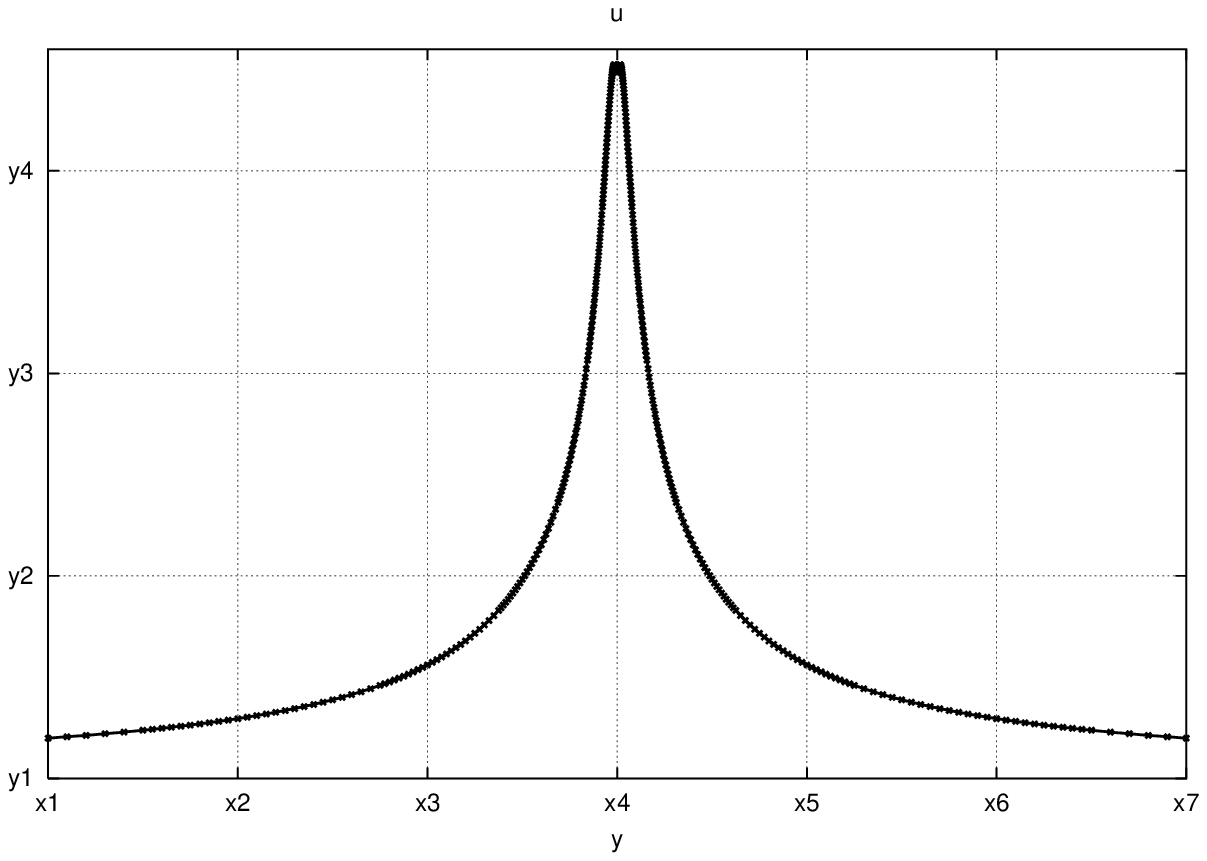,scale=0.6}}

        \psfrag{u}[c][r]{{$u$}}
        \psfrag{y}[c][t]{{$\begin{array}{cc} \\ y/M \\[1mm] (x=3M,\,z=0)\end{array}$}}
        \psfrag{x1}[t][r]{{\small $-0.3$}}
        \psfrag{x2}[t][r]{{\small $-0.15$}}
        \psfrag{x3}[t][r]{{\small $$}}
        \psfrag{x4}[t][r]{{\small $0.15$}}
        \psfrag{x5}[t][r]{{\small $0.3$}}
        \psfrag{y1}[r][t]{{\small $0.016$}}
        \psfrag{y2}[r][r]{{\small $0.017$}}
        \psfrag{y3}[r][r]{{\small $0.018$}}

        \put(8,0.5){\epsfig{file=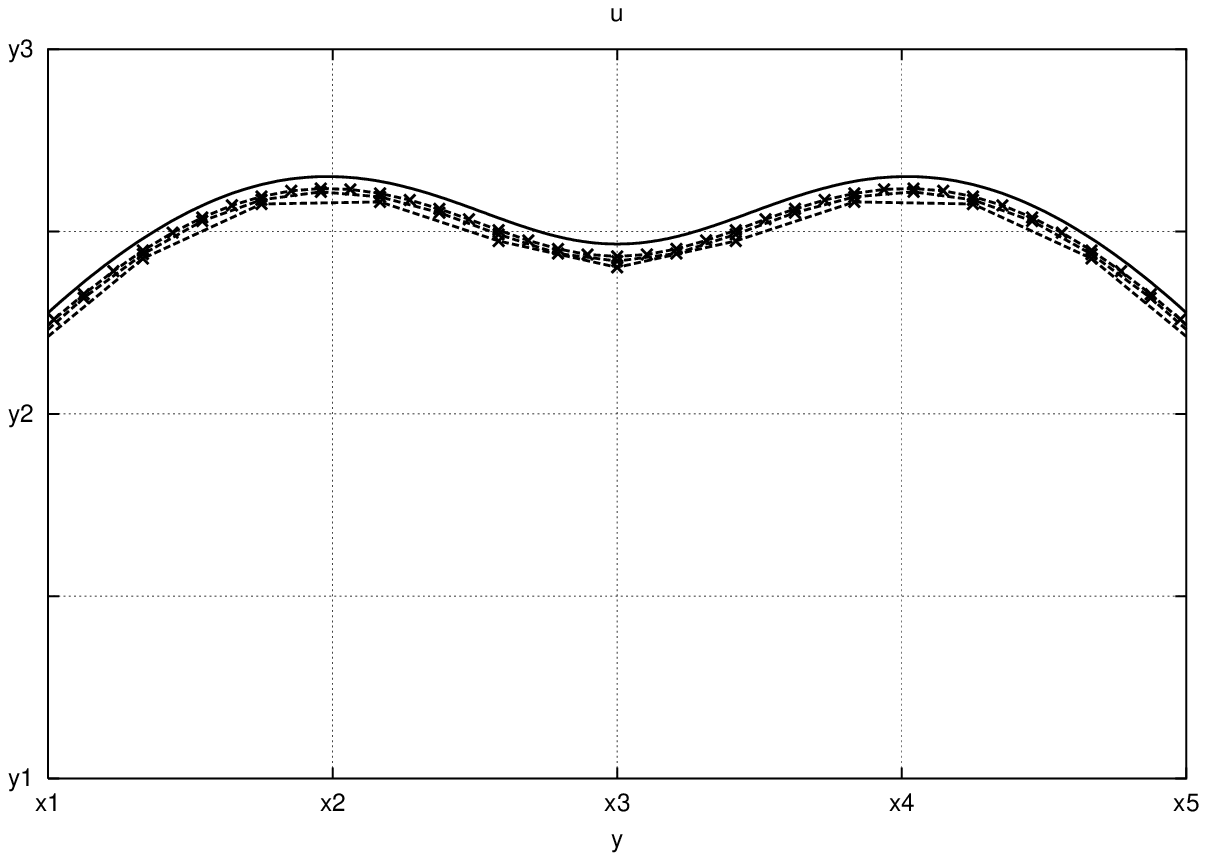,scale=0.6}}
        \end{picture}
\caption{Example for a solution to the Hamiltonian constraint
obtained with the spectral method and with a multigrid method on
nested Cartesian grids.  Shown is the regular part $u$ of the
conformal factor for two punctures without spin and vanishing
total linear momentum, which are located on the $x$-axis at $x=\pm
3M$. Results from the multigrid method are indicated by lines with
markers. The panels on the left show the various levels of
refinement combined into one line for the highest resolution (see
text). The panels on the right show an enlargement of the region
near one of the punctures for three resolutions of the multigrid
method. In all panels the result for the single-domain spectral
method with $n_A=n_B= 40$ and $n_\varphi=20$ is shown as a solid
line without markers. Note that on this scale the methods agree
well both far away and close to the punctures.
}
\label{figbam}
\end{figure*}

In Fig.\ \ref{figbam} we compare the result for $u$ obtained by
the spectral method with $u$ computed by the second order finite
difference multigrid method on a fixed mesh refinement implemented
in BAM previously \cite{Brandt97b,Bruegmann:2003aw}. As an example we
picked the parameters $b = 3M$, $m_+ = m_- = 0.5M$, and $P_+^i = -
P_-^i = 0.2M \delta^i_2$. The ADM linear momentum at infinity
vanishes. For the purpose of this discussion we have defined
$M=m_+ + m_-$. For these parameters we can restrict the 
computational domain to one quadrant
of a Cartesian box centered at the origin. We computed the
multigrid data at three overall resolutions using 7 levels of
refinement with approximately the same geometrical layout of the
boxes. The highest resolution was obtained for
$98\times98\times50$ points on the finest level,
while to test convergence we successively doubled the grid spacing, 
resulting in a grid spacing of $M/64$,
$M/32$, and $M/16$ on the finest level. The face of the outermost
box is located at about $48M$ in each case.

The main result is that the two methods verify each other quite
accurately on this scale near the punctures and also for large $x$.
Near the punctures it is a question of resolution
whether the known feature of a local indentation is fully resolved.
For the chosen parameters both methods reach
this level of resolution, but the spectral method uses significantly
less resources.

\section{Two punctures in the test mass limit}
\label{TestMassLimit}

Apart from the high accuracy that can be achieved by spectral methods,
they also prove to be very useful for the investigation of critical
and limiting situations. For the binary black hole initial data
problem, a situation of this kind is encountered when the two
gravitational sources possess very different masses. It is the aim of
this section to apply our spectral method for the binary puncture
initial data problem in this limiting case.

As a first step we perform the test mass limit analytically. The results
arising from this study will then be compared to those obtained by the
spectral method for a small mass ratio.

We consider two non-spinning punctures with bare masses $m_-$ and
$m_+$ located on the $y$-axis at $y=0$ and $y=D$ respectively and
perform the test mass limit by choosing $m_-\to 0$ with $D$ and $m_+$
held fixed.  We moreover assume that the linear momenta are given by
($v_-$ and $\hat P_+^i$ fixed, $v_-\neq 0$)
\begin{equation}
 P^i_-=m_-v_-\delta^i_1,\quad P^i_+=m_- \hat P^i_+,
\end{equation}
which implies that the total linear momentum vanishes as $m_-\to
0$. Thus, in this limit we will have placed ourselves in a frame
in which the total linear momentum vanishes.

In order to understand the behavior of $u$ in the entire spatial
domain, we have to consider two different ways of performing this
limit separately:
\begin{enumerate}
    \item If we calculate the auxiliary function $u$ at a given spatial point at some finite
    distance from the origin, i.e.\ at {\em fixed} coordinates $(x,y,z)\neq(0,0,0)$,
    we will find that $u$ tends to zero in this limit. In particular,
    \begin{equation}
        \label{u_awayfrom_m-}
        \lim_{m_-\to 0}\left(\frac{u}{m_-}\right)=\frac{\Delta\mu_\infty}{2r} ,
    \end{equation}
    where $r=\sqrt{x^2+y^2+z^2}$. The physical meaning of the constant $\Delta\mu_\infty$
    (which is obtained through the second limiting process, see below)
    with respect to the system's relative binding energy in this limit will be discussed
    in Sec.~\ref{BindingEnergy}.
    \item If, on the contrary, we hold the {\em relative} coordinates
    $(\tilde{x},\tilde{y},\tilde{z})=(x/m_-,y/m_-,z/m_-)$ fixed, 
	we maintain finite values for $u$ at these spatial points.
    In particular, the resulting $u$ obeys a constraint equation valid for a
    modified single-puncture initial data problem with non-vanishing linear momentum
    $\boldsymbol{P}$ (and no spin). The above constant $\Delta\mu_\infty$
    can be read off from these data.
\end{enumerate}

For both ways of establishing the test mass limit,
we rewrite the Hamiltonian constraint as an integral equation,
\begin{equation}
  u(\boldsymbol{x})=\frac{1}{32\pi}\int\limits_{\mathbb{R}^3}
    \frac{\psi^{5}K_{ij}K^{ij}}
    {|\boldsymbol{x}-\boldsymbol{x}\,'|}d^3\boldsymbol{x}\,'.
\end{equation}
Introducing spherical coordinates
\begin{eqnarray}
  x&=&r\cos\vartheta, \nonumber \\
  y&=&r\sin\vartheta\cos\varphi, \\
  z&=&r\sin\vartheta\sin\varphi, \nonumber
\end{eqnarray}
this integral equation becomes
\begin{widetext}
\begin{eqnarray}
  u(r,\vartheta,\varphi)&=&\frac{m_-^2v_-^2}{32\pi}
  \int_0^{2\pi}d\varphi'\int_0^\pi \sin\vartheta' d\vartheta'\int_0^\infty
  \frac{r'^2dr'}{|\boldsymbol{x}-\boldsymbol{x}\,'|}\times
\nonumber \\
& & \times
  \left[
    \left(1+\frac{m_+}{2r_+}+\frac{m_-}{2r'}+u\right)^{-7}
    \left(\frac{9}{2r'^4}(1+2\cos^2\vartheta')
        +\frac{g}{r'^2r_+^2}+\frac{h}{r_+^4}\right)
  \right] ,
\end{eqnarray}
\end{widetext}
where $\boldsymbol{x}\,'=(x',y',z')^T$ with
\begin{eqnarray}
  \nonumber x'&=&r'\sin\vartheta', \\
  y'&=&r'\cos\vartheta'\cos\varphi', \\
  \nonumber z'&=&r'\cos\vartheta'\sin\varphi'\,,
\end{eqnarray}
and
\begin{equation} r_+=\sqrt{x'^2+(y'-D)^2+z'^2}.\end{equation}
The functions $g$ and $h$ depend on $\hat P^i_+$, $v_-$ and $D$.
They remain finite everywhere.

We now perform the two different limits:
    
	1. Consider fixed values $r>0$. We split the
    integration with respect to $r'$ such that (a)
    $r'\in[r/2,\infty)$ and (b) $r'\in[0,r/2]$.

    For (a) observe that for $r'\geq r/2$ the term
    \begin{eqnarray*}
        &&\left[
            \left(1+\frac{m_+}{2r_+}+\frac{m_-}{2r'}+u\right)^{-7}\times\right. \\
            &&\;\left.\left(\frac{9}{2r'^4}(1+2\cos^2\vartheta')
            +\frac{g}{r'^2r_+^2}+\frac{h}{r_+^4}\right)
            \right]
    \end{eqnarray*}
    remains regular in the limit $m_-\to 0$,
    and thus, the contribution of the corresponding Poisson integral, evaluated
    for $r'\in[r/2,\infty)$, is of order ${\cal O}(m_-^2)$.

    Performing for the remaining near-zone integral (b) the substitution
    $r'=m_- s'$ leads us to
    \begin{eqnarray*}
        &&  \frac{18m_-v_-^2}{\pi r}
            \int_0^{2\pi}d\varphi'\int_0^\pi \sin\vartheta' d\vartheta'\int_0^{r/(2m_-)}
            ds'\times \\
        &&\qquad\left(s'^5\,\frac{1+2\cos^2\vartheta'+{\cal O}(s'm_-)}
                                    {[1+2s'(1+m_+/(2D)+u)]^7}\right) ,
    \end{eqnarray*}
    from which it follows that
    \begin{equation}
    \label{u_away_Test}
    \lim_{m_-\to 0}\frac{u(r,\vartheta,\varphi)}{m_-}
        =\frac{\Delta\mu_\infty}{2r}
    \end{equation}
    with the constant $\Delta\mu_\infty$ given by
    \begin{equation}
        \Delta\mu_\infty =\frac{36v_-^2}{\pi}\left\{
        \begin{array}{ll}
            \int\limits_0^{2\pi}d\varphi\int\limits_0^\pi
            \sin\vartheta(1+2\cos^2\vartheta)d\vartheta\int\limits_0^\infty ds\;\times \\[5mm]
            s^5\,\left[1+2s(1+m_+/(2D)+\tilde{u})\right]^{-7} .
        \end{array}
        \right.
    \end{equation}
    Here, the function $\tilde{u}$ is defined by
    \begin{equation}
        \tilde{u}(s,\vartheta,\varphi)=\lim_{m_-\to 0}u(m_-s,\vartheta,\varphi),
    \end{equation}
    and turns out to be the auxiliary potential resulting from the second limit,
    which we will discuss now.

    2. Take for the fixed relative distance limit $r=m_-s$ with $s$ fixed, $s\geq 0$, for
    which we may perform the analogous steps as in the previous case.
    We calculate the first integral for $r'\in[D/2,\infty)$, and
    again get only a contribution of order ${\cal O}(m_-^2)$. For
    the near-zone integral we obtain
    \begin{eqnarray*}
        &&  \frac{18v_-^2}{\pi}
            \int_0^{2\pi}d\varphi'\int_0^\pi \sin\vartheta' d\vartheta'\int_0^{D/(2m_-)}
            \frac{ds'}{|\boldsymbol{\,\tilde{x}}-\boldsymbol{\tilde{x}}\,'|}\times\\
        &&\qquad\left(s'^5\frac{1+2\cos^2\vartheta'+{\cal O}(s'm_-)}
                                {[1+2s'(1+m_+/(2D)+u)]^7} \right)
    \end{eqnarray*}
    with vectors 
	\begin{eqnarray*}
		\begin{array}{lcccl}
			\boldsymbol{\tilde{x}}&=&(\tilde{x},\tilde{y},\tilde{z})^T&=&\boldsymbol{x}/m_-,  \\
			\boldsymbol{\tilde{x}}\,'&=&(\tilde{x}',\tilde{y}',\tilde{z}')^T&=&\boldsymbol{x}'/m_-,
		\end{array}
	\end{eqnarray*}
    where
    \begin{eqnarray*}\qquad
		\begin{array}{lcllcl}
			\tilde{x}&=&s\sin\vartheta,\quad			& \tilde{x}'&=&s'\sin\vartheta',\\
			\tilde{y}&=&s\cos\vartheta\cos\varphi,\quad & \tilde{y}'&=&s'\cos\vartheta'\cos\varphi',\\
			\tilde{z}&=&s\cos\vartheta\sin\varphi,\quad & \tilde{z}'&=&s'\cos\vartheta'\sin\varphi'.
		\end{array}
    \end{eqnarray*}
    This leads in the limit $m_-\to 0$ to an integral equation for
    the function $\tilde{u}$
    introduced above. Equivalently, we may consider the corresponding
    differential equation
    \begin{equation}
    \label{Ham_TestPunc}
    \triangle\tilde{u}+\frac{9v_-^2}{16s^4}\tilde{\psi}^{-7}(1+2\cos^2\vartheta)=0
    \end{equation}
    with
    \begin{equation}
    \label{Psi_Test}
        \tilde{\psi}=1+\frac{m_+}{2D}+\frac{1}{2s}+\tilde{u},
    \end{equation}
    and the Laplace operator taken in the spherical coordinates
    $(s,\vartheta,\varphi)$. In particular it follows that
    \begin{equation}
    \label{Del_mu_def}
        \lim_{s\to\infty}2s\tilde{u}(s,\vartheta,\varphi)=\Delta \mu_\infty.
    \end{equation}
    We moreover see that for the function
    \begin{equation}
        \hat{u}=\hat{m}\tilde{u}\quad\mbox{with}\quad \hat{m}=\left(1+\frac{m_+}{2D}\right)^{-1}
    \end{equation}
    we recover the equation valid for a single-puncture initial data problem (without spin),
    see Sec.~\ref{onepuncwithP}.
        The (dimensionless) bare mass is just $\hat{m}$, and the
    corresponding momentum reads
    \begin{equation}
        \pi_x=v_-\hat{m}^4.
    \end{equation}
The above analytic study shows that we can use our spectral
methods applied to a single puncture with non-vanishing momentum
(as described in Sec.~\ref{onepuncwithP}) in order to evaluate the
test mass limit with algebraic convergence of fourth order. These
results can be compared with the values obtained for a
corresponding two puncture initial data problem with a small mass
ratio, see Table 1. In this table one finds the value
$u_-=u(0,0,0)$ at the origin (i.e.\ at the `light' puncture), the
expression
\begin{equation} \frac{2D}{m_-} u_+=\frac{2D}{m_-} u(0,D,0)\end{equation}
(i.e.\ at the `heavy' puncture), and the limit
\begin{equation} \lim_{r\to\infty}\left(\frac{2ru}{m_-}\right) , \end{equation}
where the latter two tend to $\Delta\mu_\infty$ as $m_-\to 0$.
We have chosen a particular example where the distance $D$ and the
velocity $v_-$ obey the relations valid for the last stable circular
orbit of a test particle in the gravitational field of a Schwarzschild
black hole of mass $m_+$:
\begin{equation}
    \label{test_mass_example}
    \frac{D}{m_+} = \frac{5}{2}+\sqrt{6}\;,\quad v_-=\frac{4\sqrt{3}}{5+2\sqrt{6}}.
\end{equation}
Moreover, we simply set $\hat P^i_+=-v_-\delta^i_1$.

It turns out that for ratios $m_-/m_+\geq 10^{-3}$ the spectral
scheme yields reliable results that approach those of the test
mass limit. For mass ratios of $10^{-3}$, four digits of accuracy
are obtained for the given order of approximation from the
two-puncture calculation, while six
digits are obtained with the single-puncture method
for the test mass limit.

\begin{center}
    \begin{table}
        \begin{tabular}{|c|c|c|c|}
            \hline
            $m_-/m_+$   & $u_-$     & $2Du_+/m_-$   & $\lim\limits_{r\to\infty}(2ru/m_-)$ \\[2mm]
            \hline
            $10^{-1}$           & 0.03417   & 0.2011                & 0.1688 \\
            $10^{-2}$           & 0.03406   & 0.1635                & 0.1601 \\
            $10^{-3}$           & 0.03406   & 0.1596                & 0.1592 \\
            \hline
            $0$                 & 0.0340568 & 0.159094              & 0.159094\\
            \hline
        \end{tabular}
        \caption{Test mass limit $m_-\to 0$ for the
        representative example with values given in
        (\ref{test_mass_example}) with
        $P_-^i=-P^i_+=-m_-v_-\delta^i_1$. For the above
        non-vanishing mass ratios we used the spectral method
        for the binary-puncture initial data problem with
        $n_A=n_B=2n_\varphi=100$. The last line has been
        calculated with the spectral method for the single-puncture
        initial data problem with $n_A=n_B=70$,
        $n_\varphi=4$.}
\end{table}
\end{center}


\section{Binding energy in the test mass limit}
\label{BindingEnergy}

In this section we use the results of the previous section to
compute the binding energy of two punctures without spin in the
limit of vanishing mass ratio. The aim will be to compare the
binding energy in this test mass limit with the binding energy of
a test particle in Schwarzschild spacetime. The deviation of the
puncture binding energy from the Schwarzschild result will yield a
quantitative statement about how realistic puncture data are in
this limit. If punctures were completely realistic we should
recover the Schwarzschild results. A related study of small mass
ratios (up to $1/32$) has already been performed by Pfeiffer
\cite{Pfeiffer:thesis} for excision-type initial data with
Bowen-York extrinsic curvature, and also to a limited extent for
conformal thin sandwich initial data.

In order to define a binding energy we need a notion of the total
mass as well as of the local black hole masses. 
The ADM mass at infinity yields a well-defined
global mass. For two punctures it is given by
\begin{equation}
M^{ADM}_{\infty} = m_+ + m_- + \Delta M_{\infty} ,
\end{equation}
where
\begin{equation}
\Delta M_{\infty} = -\frac{1}{2\pi} \oint_{\infty} \nabla_i u \ dA^i
 = \lim_{r\rightarrow \infty} 2r u.
\end{equation}
On the other hand, it is impossible to unambiguously define local black
hole masses in general. In the following we choose the ADM mass
\begin{equation}
\label{local_Mass}
M^{ADM}_{\pm} = (1+u_{\pm}) m_{\pm} + \frac{m_+ m_-}{2D}
\end{equation}
computed in the asymptotically flat region at each puncture
as a measure of the local black hole mass  \cite{Tichy03a}.
Here $u_{+}$ and $u_-$ are the values of $u$ at each puncture.
As shown by Beig \cite{Beig:2000ei}, this definition of local mass has
the following advantage. For a single slowly moving puncture with momentum
$\boldsymbol{P}^{ADM}_{\infty} = \boldsymbol{P}_-$,
the ADM energy $E^{ADM}_{\infty}$ at infinity
is related to the ADM mass $M^{ADM}_{-}$ as measured
in the asymptotically flat region near the puncture by
\begin{equation}
E^{ADM}_{\infty} = M^{ADM}_{-}
+ \frac{ (\boldsymbol{P}^{ADM}_{\infty})^2 }{ 2 M^{ADM}_{-} }
+ O\left( (\boldsymbol{P}^{ADM}_{\infty})^4 \right) ,
\end{equation}
which is just what one expects if the local mass definition is
reasonable. If, for example, one uses instead the bare mass $m_-$
as the definition of local mass, one finds (e.g. \cite{Beig:2000ei,Baker02a})
\begin{equation}
E^{ADM}_{\infty} = m_-
+ \frac{5}{8}\frac{(\boldsymbol{P}^{ADM}_{\infty})^2}{m_-} +
                    O\left( (\boldsymbol{P}^{ADM}_{\infty})^4 \right),
\end{equation}
which is incompatible with special relativity.

Next, we define the binding energy for two punctures by
\begin{eqnarray}
\label{E_b_def}
E_b &=& M^{ADM}_{\infty} - M^{ADM}_{+} - M^{ADM}_{-}    \nonumber \\
    &=& \Delta M_{\infty} - m_+ u_+ - m_- u_- - \frac{m_+ m_-}{D}.
\end{eqnarray}
In the test mass limit of $m_-\to 0$, it follows from Eq.~(\ref{u_awayfrom_m-}) that
\begin{equation}
\label{DelM_Delmu}
\lim_{m_-\rightarrow 0} \Delta M_{\infty}/m_- = \Delta \mu_{\infty}
\end{equation}
and
\begin{equation}
\lim_{m_-\rightarrow 0} u_+ = 0.
\end{equation}
Thus $E_b$ goes to zero in this limit. We therefore consider $E_b/\mu$, where
\begin{equation}
\mu = M^{ADM}_{+} M^{ADM}_{-} / \left(M^{ADM}_{+} + M^{ADM}_{-}\right)
\label{mu}
\end{equation}
is the reduced mass. With the help of
Eqs.~(\ref{E_b_def}),
(\ref{u_awayfrom_m-}), (\ref{Del_mu_def}) and (\ref{DelM_Delmu})
we find that
\begin{eqnarray}
\lim_{m_- \to 0} \frac{E_b}{\mu}
&=& \lim_{m_- \to 0}
    E_b \left(\frac{1}{M^{ADM}_{+}} + \frac{1}{M^{ADM}_{-}}\right)\nonumber \\
&=& \frac{ \Delta \mu_{\infty} \left( 1 - \frac{m_+}{2D} \right)
          - u_- - \frac{m_+}{D}     }
         { 1 + u_- + \frac{m_+}{2D} } .
\end{eqnarray}
This binding energy can now be compared with the binding energy of a test
particle in Schwarzschild spacetime. For circular geodesics in
Schwarzschild the binding energy, angular momentum and angular velocity
observed at infinity are given by
\begin{equation}
\frac{E_{b,S}}{\mu}=\frac{ (2r-M)^2 }{(2r+M) \sqrt{ 4r^2 -8M r +M^2 } } - 1 ,
\end{equation}
\begin{equation}
\frac{L_S}{\mu}=\frac{(2r+M)^2 }{2r^2}
                \sqrt{ \frac{ M r^3 }{(4r^2 -8 M r +M^2)} } ,
\end{equation}
and
\begin{equation}
\label{Omega_S}
\Omega_S = \sqrt{ \frac{64 M r^3}{(2r+M)^6} } ,
\end{equation}
respectively, where $M$ is the mass of the Schwarzschild black hole, $\mu$
is the mass of the test particle and $r$ is the orbital radius in isotropic
coordinates.

In order to compute $E_b/\mu$ for punctures in the test mass limit
we have to solve Eq.~(\ref{Ham_TestPunc}) with the appropriate
velocity $v_-$ for circular orbits. This raises two questions. The
first is how to choose the coordinate distance $D$ between the two
punctures if one wants to compare with a test particle in
Schwarzschild at isotropic radial coordinate $r$. The answer is
that in the limit of $m_- \to 0 $ the spacetime is determined by
the puncture with bare mass $m_+$, so that one simply obtains
Schwarzschild in isotropic coordinates, which allows us to set
\begin{equation}
D = r .
\end{equation}
The second question is how one should choose $v_-$ for two
punctures in circular orbit. One could, for example, obtain $v_-$
by requiring equality of Komar and ADM mass, which is a necessary
condition for the existence of a helical Killing vector, as done
in \cite{Tichy03b}. An alternative would be the effective potential
method \cite{Baumgarte00a}. Each of these methods will give
a binding energy and an angular momentum for the so-defined
circular orbits and in general we do not expect the binding energy
and angular momentum to exactly agree with the Schwarzschild
results. For simplicity and in order to eliminate possible errors
in the angular momentum we choose
\begin{equation}
v_- = \frac{L_S/\mu}{ r}
\end{equation}
so that the angular momentum of the light puncture exactly
equals the angular momentum of a test particle in Schwarzschild spacetime.
\begin{figure}
\epsfxsize=8.5cm \epsfbox{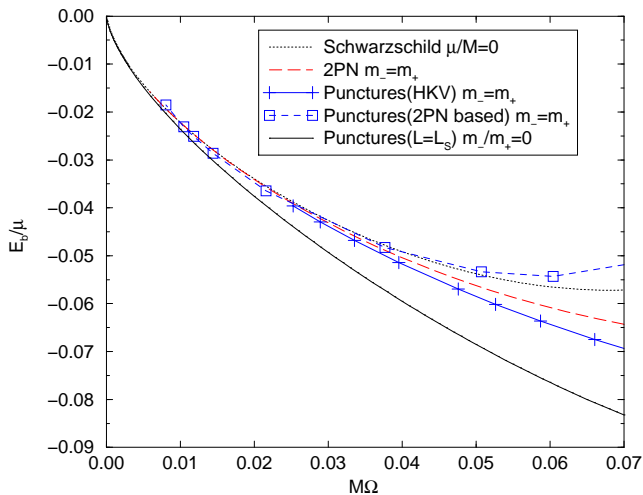} \vspace{.4cm} \caption{ The
solid line on the bottom shows the binding energy versus angular
velocity for two punctures in circular orbit in the test mass
limit. For comparison we also show the binding energy of a test
particle in Schwarzschild (dotted line). In addition we show
several binding energies for circular orbits in the equal mass
case. The post-2-Newtonian binding energy (broken line) is close
to puncture data based on an approximate helical Killing vector
(pluses) as well as to puncture data based on post-2-Newtonian
data (squares), and also to the Schwarzschild result. }
\label{Ebind_vs_Omega}
\end{figure}

Using our spectral method with $n_A = n_B = 70$ and $n_{\varphi} = 4 $
we have computed the binding energy for punctures in the test mass
limit. The result is plotted in Fig.~\ref{Ebind_vs_Omega} versus the
angular velocity given in Eq.~(\ref{Omega_S}). Also shown are the
results for circular orbits in Schwarzschild and several other binding
energies in the equal mass case taken from \cite{Tichy02} and
\cite{Tichy03b}. One can see that the binding energy for punctures
(solid line on bottom) in the test mass limit does not agree with the
binding energy of a test particle in Schwarzschild (dotted line),
except in the Newtonian limit of small $M\Omega$. The discrepancy
reaches about $50\%$ at the innermost stable circular orbit (ISCO) of
Schwarzschild, which means that the amount of energy radiated before
reaching the Schwarzschild ISCO is too large by $50\%$ and
that the location of the ISCO predicted by puncture data is wrong.
This means that, for the assumptions made in the definition of the
binding energy, puncture data are not realistic for extreme mass
ratios and that one cannot expect to obtain reliable
predictions about the gravitational waves emitted.

Let us point out two possible reasons for the discrepancy. One is that
it is not clear whether there are alternatives to our definition of mass,
(\ref{local_Mass}), that change the result. Another issue is that
it is known that there is `artificial' radiation present in puncture
data. Such radiation could contribute at the observed level to the
ratio of infinitesimal binding energy to infinitesimal mass.

Interestingly, the curve for puncture data in Fig.\
\ref{Ebind_vs_Omega} in the equal mass case (marked by pluses) is
much closer to both the Schwarzschild (dotted line) case and the
post-2-Newtonian (broken line) results, as well as to the results
of the numerical method based on post-2-Newtonian data (marked by
squares) discussed in \cite{Tichy02}. This might indicate that
artificial radiation affects the binding energy per reduced mass
for comparable mass puncture data less than in the test mass
limit.

Note also that in \cite{Faye:2003sw} a method has been described in which
conformally flat black hole data does indeed lead to the correct
Schwarzschild result for the binding energy in the test mass
limit. That method is quite different, for example $u$ is approximated
by zero and the local masses entering the binding energy are defined
differently. At this point it is not clear how to make contact with
our approach, but this is clearly an important question for future
research.

\section{Conclusion}
\label{Conclusion}

In Cartesian coordinates the regular part of the conformal factor
of puncture initial data is only ${\cal C}^2$-differentiable at
the punctures. Therefore, a numerical implementation based on a
spectral method is expected to be at most fourth-order
algebraically convergent. However, one can overcome this problem
by introducing appropriate coordinates in which the solution is
smooth at the punctures. 
In particular, our transformation maps the entire
$\mathbb{R}^3$ onto a single rectangular domain with the punctures
at the boundary.

We have demonstrated rapid convergence of our single-domain spectral
method and obtained highly accurate numerical solutions. Moreover, we
have provided a comparison to a numerical implementation with finite
differences in Cartesian coordinates and found good agreement.

While our coordinate transformation renders puncture data smooth at
the punctures, in general the fall-off of the extrinsic curvature
appears to imply the existence of logarithmic terms such that the
solution is only ${\cal C}^4$ at infinity if the total linear momentum
vanishes, and only ${\cal C}^2$ otherwise. This behavior is a
consequence of the fall-off of the Bowen-York extrinsic curvature and
as such unrelated to the puncture construction. It is an interesting
but to our knowledge mostly open question which other approaches to
construct initial data for black holes share or avoid the problem of
logarithmic terms at infinity.

As an application of our spectral method for punctures, we have
considered small mass ratios, and the corresponding results approach
the test mass limit which was obtained through a semi-analytic
limiting procedure. Finally, we have computed the binding energy of
two punctures in the test mass limit and compared it to the binding
energy of a test particle in Schwarzschild spacetime and to binding
energies in the equal mass case. We find that in the test mass limit
the binding energy per mass deviates from the Schwarzschild result by
about $50\%$ at the Schwarzschild ISCO, while the binding energy of
two punctures in the equal mass case is close to post-Newtonian
results, if the ADM mass at each puncture is used to define the local
black hole masses. This should be compared with \cite{Faye:2003sw},
where by a different method conformally flat black hole data does lead
to the proper test mass limit.

The study of specific coordinate transformations might also help in
reducing the number of domains that are used by methods for binary
black hole excision data.  We have started a corresponding
investigation based on a coordinate transformation that requires two
coordinate patches, and we intend to apply spectral methods. Within
the analysis of these data we plan among other things to go into the
matter of possible logarithmic expansion terms of the conformal factor
in the context of binary black hole excision data.

\begin{acknowledgments}
It is a pleasure to thank P. Laguna and U. Sperhake for discussions.
We are grateful to S. Dain for pointing us to the logarithmic 
expansion terms of the conformal factor at higher orders.
This work was supported by NSF grants PHY-02-18750 and PHY-02-44788.
We also acknowledge the support of the Center for Gravitational Wave
Physics funded by the National Science Foundation under Cooperative
Agreement PHY-01-14375. M. A. was supported by DFG stipend AN 384/1-1.
\end{acknowledgments}


\bibliography{references}

\end{document}